\documentclass[journal]{IEEEtran}

\usepackage{url,cite,epsf,xspace,amsmath,amssymb,amsfonts,graphicx,subfigure,color,setspace}
\usepackage{multirow}

\def\bf{\mathbf}

\usepackage{mathrsfs} 
\usepackage{times,amstext,amsmath,amssymb,latexsym,shapepar,array,epsfig,bm,paralist}
\usepackage{amsfonts,subfigure}
\usepackage{amssymb}
\usepackage{amsthm,amsmath}
\usepackage{graphicx}
\graphicspath{{figs/}}

\usepackage{multirow}

\usepackage{algorithm}
\usepackage{algorithmic}
\usepackage{bm}

\providecommand{\norm}[1]{\left \lVert#1 \right  \rVert}

\providecommand{\mb}[1]{\boldsymbol{#1}}

\providecommand{\mhb}[1]{\hat{\boldsymbol{#1}}}

\newcommand{\Dmat}{{\bf D}}
\newcommand{\Emat}{{\bf E}}

\newcommand{\Imat}{{\bf I}}

\newcommand{\Smat}{{\bf S}}

\newcommand{\Xmat}{{\bf X}}

\newcommand{\bv}{\boldsymbol{b}}

\newcommand{\dv}{\boldsymbol{d}}
\newcommand{\ev}{\boldsymbol{e}}

\newcommand{\pv}{\boldsymbol{p}}

\newcommand{\sv}{\boldsymbol{s}}

\newcommand{\Lambdamat}{\boldsymbol{\Lambda}}

\newcommand{\Sigmamat}{\boldsymbol{\Sigma}}

\newcommand{\Omegamat}{\boldsymbol{\Omega}}

\newcommand{\lambdav}{\boldsymbol{\lambda}}
\newcommand{\muv}{\boldsymbol{\mu}}

\newcommand{\piv}{\boldsymbol{\pi}}

\newcommand{\phiv}{\boldsymbol{\phi}}

\newtheorem{thm}{Theorem}[section]

\newtheorem{lem}[thm]{Lemma}

\newcommand{\beq}{\begin{equation}}
\newcommand{\eeq}{\end{equation}}
\newcommand{\beqs}{\begin{eqnarray}}
\newcommand{\eeqs}{\end{eqnarray}}
\newcommand{\barr}{\begin{array}}
\newcommand{\earr}{\end{array}}

\ifCLASSINFOpdf
\else
\fi
\usepackage[update,prepend]{epstopdf}
\newcommand{\Real}{\mathbb{R}}
\usepackage{color}

\begin{document}


%
\title{
{Multichannel Electrophysiological Spike Sorting via Joint Dictionary Learning \& Mixture Modeling}
}%
%

%

\author{David E.\ Carlson, Joshua T.\ Vogelstein, Qisong Wu, Wenzhao Lian, Mingyuan Zhou, Colin R.\ Stoetzner,  \\ Daryl Kipke, Douglas Weber,  David B.\ Dunson and Lawrence Carin
\thanks{Q.\ Wu, D.\  Carlson, J.\ T.\ Vogelstein,  W.\  Lian, M.\  Zhou and L.\  Carin are with the Department
of Electrical and Computer Engineering, Duke University, Durham, NC, USA}
\thanks{C.\ R.\  Stoetzner and D.\  Kipke are with the Department of Biomedical Engineering, University of Michigan, Ann Arbor, MI, USA}
\thanks{D.\  Weber is with the Department of Biomedical Engineering, University of Pittsburgh, Pittsburgh, PA, USA}
\thanks{J.\ T.\ Vogelstein and D.\  Dunson are with the Department of Statistical Science, Duke University, Durham, NC, USA}
\thanks{Manuscript received October 27, 2012.}}

%
%

\markboth{IEEE Transactions on Biomedical Engineering}%
{Shell \MakeLowercase{\textit{et al.}}: Bare Demo of IEEEtran.cls for Journals}
%



\maketitle

\begin{abstract}

We propose a methodology for joint feature learning and clustering of multichannel extracellular electrophysiological data, across multiple recording periods for action potential detection and classification (``spike sorting'').
Our methodology improves over the previous state of the art principally in four ways.  First, via sharing information across channels, we can better distinguish between single-unit spikes and artifacts.  Second, our proposed ``focused mixture model'' (FMM) deals with units appearing, disappearing, or \emph{reappearing} over multiple recording days, an important consideration for any chronic experiment.  Third, by jointly learning features and clusters, we improve performance over previous attempts that proceeded via a two-stage learning process.  
Fourth, by directly modeling spike rate, we improve detection of sparsely firing neurons. Moreover, our Bayesian methodology seamlessly handles missing data.  
We present state-of-the-art performance without requiring manually tuning hyperparameters, considering both a public dataset with partial ground truth and a new experimental dataset.\footnote{Copyright (c) 2013 IEEE. Personal use of this material is permitted. However, permission to use this material for any other purposes must be obtained from the IEEE by sending an email to pubs-permissions@ieee.org.}  
\end{abstract}


%
\IEEEpeerreviewmaketitle

\section{Introduction} \label{sec:intro}
%
%
%
%

\IEEEPARstart{S}{pike} sorting of extracellular electrophysiological data is an important problem in contemporary neuroscience, with applications ranging from brain-machine interfaces \cite{Nicolelis2009} to neural coding \cite{Rieke1997} and beyond.  Despite a rich history of work in this area \cite{Wheeler1991, Einevoll2012}, room for improvement remains for automatic methods. In particular, we are interested in sorting spikes from multichannel longitudinal data, where longitudinal data potentially consists of many experiments conducted in the same animal over weeks or months.  


Here we propose a Bayesian generative model and associated inference procedure. Perhaps the most important advance in our present work over previous art is our joint feature learning and clustering strategy.  More specifically, 
%
standard pipelines for processing extracellular electrophysiology data consist of the following steps:
($i$) filter the raw sensor readings, 
($ii$) perform thresholding to ``detect" the spikes, 
($iii$) map each detected spike to a feature vector, and then
($iv$)  cluster the feature vectors
\cite{Lewicki}. 
Our primary conceptual contribution to spike sorting methodologies is a novel unification of steps ($iii$) and ($iv$) that utilizes all available data in such a way as to satisfy all of the the above criteria. This \emph{joint} dictionary learning and clustering approach improves results even for a single channel and a single recording experiment ($i.e.$, not longitudinal data).  Additional localized recording channels improve the performance of our methodology by incorporating more information.  More recordings allow us to track dynamics of firing over time.
%

Although a comprehensive survey of previous spike sorting methods is beyond the scope of this manuscript, below we provide a summary of previous work as relevant to the above listed goals.
Perhaps those methods that are most similar to ours include a number of recent Bayesian methods for spike sorting \cite{Wood2009,Bo2011}.  One can think of our method as a direct extension of theirs with a number of enhancements. Most importantly, we learn features for clustering, rather than simply using principal components. We also incorporate multiple electrodes,  assume a more appropriate prior over the number of clusters, and address longitudinal data.


Other popular methods utilize principal components analysis (PCA) \cite{Lewicki} or wavelets \cite{Letelier2000} to find low-dimensional representations of waveforms for subsequent clustering.  These methods typically require some manual tuning, for example, to choose the number of retained principal components.  Moreover, these methods do not naturally handle missing data well. Finally, these methods choose low-dimensional embeddings for reconstruction and are not necessarily appropriate for downstream clustering. 


Calabrese \emph{et al.} \cite{Calabrese2010} recently proposed a Mixture of Kalman Filters (MoK) model to explicitly deal with slow changes in waveform shape.  This approach also models spike rate (and even refractory period), but it does not address our other desiderata, perhaps most importantly, utilizing multiple electrodes or longitudinal data. It would be interesting to extend that work to utilize learned time-varying dictionaries rather than principal components.

Finally, several recently proposed methods address sparsely firing neurons \cite{Pedreira2012, Adamos2012}.  By directly incorporating firing rate into our model and inference algorithm (see Section \ref{sec:mixture}), our approach outperforms previous methods even in the absence of manual tuning (see Section \ref{sec:sparse}).

The remainder of the manuscript is organized as follows.  Section \ref{sec:models} begins with a conceptual description of our model followed by mathematical details and experimental methods for new data. Section \ref{sec:results} begins by comparing the performance of our approach to several other previous state-of-the-art methods, and then highlights the utility of a number of additional features that our method includes.  Section \ref{sec:conclusions}  summarizes and provides some potential future directions.  The Appendix provides details of the relationships between our method and other related Bayesian models or methodologies.

\section{Models and Analysis\label{sec:models}}

\subsection{Model Concept} 
\label{sub:concept1}

Our generative model derives from knowledge of the properties of electrophysiology signals.  
Specifically, we assume that each waveform can be represented as a sparse superposition of several dictionary elements, or features.  Rather than presupposing a particular form of those features ($e.g.$, wavelets), we \emph{learn} features from the data.  Importantly, we learn these features for the specific task at hand: spike sorting ($i.e.$, clustering).  This is in contrast to other popular feature learning approaches, such as principal component analysis (PCA) or independent component analysis (ICA), which learn features to optimize a different objective function (for example, minimizing reconstruction error). Dictionary learning has been demonstrated as a powerful idea, with demonstrably good performance in a number of applications \cite{Zhou2012}.  Moreover, statistical guarantees associated with such approaches are beginning to be understood \cite{Spielman2012}.  Section \ref{sec:dict} provides mathematical details for our Bayesian dictionary learning assumptions.

We \emph{jointly} perform dictionary learning and clustering for analysis of multiple spikes.
The generative model requires a prior on the number of clusters.  Regardless of the number of putative spikes detected, the number of different single units one could conceivably discriminate from a single electrode is upper bounded due to the conductive properties of the tissue.  Thus, it is undesirable to employ Bayesian nonparametric methods \cite{Antoniak74} that enable the number of clusters (each cluster associated with a single-unit event) to increase in an unbounded manner as the number of threshold crossings increases. We develop a new prior to address this issue, which we refer to as a ``focused mixture model'' (FMM). The proposed prior is also appropriate for chronic recordings, in which single units may appear for a subset of the recording days, but also disappear and reappear intermittently. Sections \ref{sec:mixture} and \ref{sec:focused} provide mathematical details for the general mixture modeling case, and our specific focused mixture model assumptions.

We are also interested in multichannel recordings.  When we have multiple channels that are within close proximity to one another, we can ``borrow statistical strength'' across the channels to improve clustering accuracy.  Moreover, we can ascertain that certain movement or other artifacts -- which would appear to be spikes if only observing a single channel -- are clearly not spikes from a single neuron, as evidenced by the fact that they are observed simultaneously across all the channels, which is implausible for a single neuron. While it is possible that different neurons may fire simultaneously and be observed coincidently across multiple sensor channels, we have found that this type of observed data are more likely associated with animal motion, and artifacts from the recording setup (based on recorded video of the animal). We employ the multiple-channel analysis to distinguish single-neuron events from artifacts due to animal movement (inferred based on the electrophysiological data alone, without having to view all of the data). 

Finally, we explicitly model the spike rate of each cluster.  This can help address refractory issues, and perhaps more importantly, enables us to detect sparsely firing neurons with high accuracy.

Because our model is fully Bayesian, we can readily impute missing data.  Moreover, by placing relatively diffuse but informed hyperpriors on our model, our approach does not require any manual tuning. And by reformulating our priors, we can derive (local) conjugacy which admits efficient Gibbs sampling.  Section \ref{sec:computations} provides details on these computations. In some settings a neuroscientist may want to tune some parameters, to tests hypotheses and impose prior knowledge about the experiment; we also show how this may be done in Section \ref{sec:tuning}.



\subsection{Bayesian dictionary learning\label{sec:dict}}

Consider electrophysiological data measured over a prescribed time interval. Specifically, let $\Xmat_{ij}\in\mathbb{R}^{T\times N}$ represent the $j${$^{th}$} signal observed during interval $i$ {(each $j$ indexes a threshold crossing within a time interval $i$)}. The data are assumed recorded on each of $N$ channels, from an $N$-element sensor array, and there are $T$ time points associated with each detected spike waveform (the signals are aligned with respect to the peak energy of all the channels). In tetrode arrays \cite{tetrode}, and related devices like those considered below, a single-unit event (action potential of a neuron) may be recorded on multiple adjacent channels, and therefore it is of interest to process the $N$ signals associated with $\Xmat_{ij}$ jointly; the joint analysis of all $N$ signals is also useful for {longitudinal analysis}, discussed in Section \ref{sec:results}.

To constitute data $\Xmat_{ij}$, {we assume} that threshold-based detection (or a related method) is performed on data measured from each of the $N$ sensor channels. When a signal is detected on any of the channels, coincident data are also extracted from all $N$ channels, within a window of (discretized) length $T$ {centered at the spikes' energy peak average over all channels}. On some of the channels data may be associated with a single-unit event, and on other channels the data may represent background noise. Both types of data (signal and noise) are modeled jointly, as discussed below.

Following \cite{Bo2011}, we employ dictionary learning to model each $\Xmat_{ij}$; however, unlike \cite{Bo2011} we jointly employ dictionary learning to all $N$ channels in $\Xmat_{ij}$ (rather than separately to each of the channels). The data are represented
\beq\Xmat_{ij}=\Dmat \Lambdamat \Smat_{ij}+\Emat_{ij},\label{eq:basic}\eeq
where $\Dmat\in\mathbb{R}^{T\times K}$ represents a dictionary with $K$ dictionary elements (columns), $\Lambdamat\in\mathbb{R}^{K\times K}$ is a diagonal matrix with sparse diagonal elements, $\Smat_{ij}\in\mathbb{R}^{K\times N}$ represents the dictionary weights (factor scores), and $\Emat_{ij}\in\mathbb{R}^{T\times N}$ represents residual/noise. Let $\Dmat=(\dv_1,\dots,\dv_K)$ and $\Emat=(\ev_1,\dots,\ev_N)$, with {$\dv_k$, $\ev_n\in\mathbb{R}^T$}. We impose priors
\beq \dv_k\sim\mathcal{N}(0,\frac{1}{T}\Imat_T)~,~~~ \ev_n\sim\mathcal{N}(0,\mbox{diag}(\eta_1^{-1},\dots,\eta_T^{-1})),\eeq
where $\Imat_T$ is the $T\times T$ dimensional identity matrix {and $\eta_t \in \Real$ for all $t$}.

We wish to impose that each column of $\Xmat_{ij}$ lives in a linear subspace, with dimension and composition to be inferred. The composition of the subspace is defined by a selected subset of the columns of $\Dmat$, and that subset is defined by the non-zero elements in the diagonal of $\Lambdamat=\mbox{diag}(\lambdav)$, with $\lambdav=(\lambda_1,\dots,\lambda_K)^T$ {and $\lambda_k \in \Real$ for all $k$}. We impose $\lambda_k\sim\nu\delta_0+(1-\nu)\mathcal{N}_+(0,\alpha_0^{-1})$, with $\nu\sim\mbox{Beta}(a_0,b_0)$ and $\delta_0$ a unit measure concentrated at zero. The hyperparameters {$a_0,b_0 \in \Real$} are set to encourage sparse $\lambdav$, and $\mathcal{N}_+(\cdot)$ represents a normal distribution truncated to be non-negative. Diffuse gamma priors are placed on $\{\eta_t\}$ and $\alpha_0$.

Concerning the model priors, the assumption $\dv_k\sim\mathcal{N}(0,\frac{1}{T}\Imat_T)$ is consistent with a conventional $\ell_2$ regularization
 on the dictionary elements. Similarly, the assumption $\ev_n\sim\mathcal{N}(0,\mbox{diag}(\eta_1^{-1},\dots,\eta_T^{-1}))$ corresponds to an $\ell_2$ fit of the data to the model, with a weighting on the norm as a function of the sample point (in time) of the signal.  We also considered using a more general noise model, with $\ev_n\sim\mathcal{N}(0,\Sigmamat)$.  These priors are typically employed in dictionary learning; see \cite{Zhou2012} for a discussion of the connection between such priors and optimization-based dictionary learning.

\subsection{Mixture modeling} \label{sec:mixture}

A mixture model is imposed for the dictionary weights $\Smat_{ij}=(\sv_{ij1},\dots,\sv_{ijN})$, with $\sv_{ijn}\in\mathbb{R}^K$; $\sv_{ijn}$ {defines} the weights on the dictionary elements for the data associated with the $n$th channel ($n$th column) in $\Xmat_{ij}$. Specifically,
\beqs & \sv_{ijn}\sim\mathcal{N}(\muv_{{z_{ij}n}},\Omegamat_{{z_{ij}n}}^{-1}),
\qquad z_{ij}\sim\sum_{m=1}^M \pi^{(i)}_m\delta_m,
\label{eq:mixture0}\\ 
&~(\muv_{{mn}},\Omegamat_{{mn}})\sim G_0(\mu_0, \beta_0,W_0, \nu_0) \label{eq:mixture}\eeqs
{where $G_0$ is a normal-Wishart distribution with
$\mu_0$ a $K$ dimension vector of zeros, $\beta_0=1$,
$W_0$ is a $K$ dimensional identity matrix, and $\nu_0=K$.
The other parameters: }$\pi^{(i)}_m>0$, $\sum_{m=1}^{M} \pi^{(i)}_m=1$, and $\{\sv_{ijn}\}_{n=1,N}$ are all associated with cluster $z_{ij}$; $z_{ij}\in\{1,\dots,M\}$ is an indicator variable defining  with which cluster $\Xmat_{ij}$ is associated{, and $M$ is a user-specified upper bound on the total number of clusters possible}.

The use of the Gaussian model in (\ref{eq:mixture0}) is convenient, as it simplifies computational inference, and the normal-Wishart distribution $G_0$ is selected because it is the conjugate prior for a normal distribution. The key novelty we wish to address in this paper concerns design of the mixture probability vector $\piv^{(i)}=(\pi_1^{(i)},\dots,\pi_{M}^{(i)})^T$.

\subsection{{Focused Mixture Model}\label{sec:focused}}

{The vector $\piv^{(i)}$ defines the probability with which each of the $M$ mixture components are employed for data recording interval $i$. We wish to place a prior probability distribution on $\piv^{(i)}$, and to infer an associated posterior distribution based upon the observed data. Let $b_m^{(i)}$ be a binary variable indicating whether interval $i$ uses mixture component $m$.  Let $\hat{\phi}_m^{(i)}$ correspond to the relative probability of including mixture component $m$ in interval $i$, which is related to the firing rate of the single-unit corresponding to this cluster during that interval.  Given this, the probability of cluster $m$ in interval $i$ is}
\beqs 
&\pi_m^{(i)}= \frac{1}{Z} b_m^{(i)}\hat{\phi}_m^{(i)} 
\label{eq:mixt}\eeqs 
{where $Z=\sum_{m^\prime=1}^M b_{m^\prime}^{(i)}\hat{\phi}_{m^\prime}^{(i)}$ is the normalizing constant to ensure that $\sum_m \pi_m^{(i)}=1$.  To finalize this parameterization, we further assume the following priors on $b_m^{(i)}$ and $\hat{\phi}_n^{(i)}$:}
\begin{multline} \label{eq:gen1}
\hat{\phi}_m^{(i)}\sim \mbox{Ga}(\phi_m,p_i/(1-p_i)), \\
\phi_m\sim\mbox{Ga}(\gamma_0,1) ,~p_i\sim\mbox{Beta}(a_0,b_0)
\end{multline}
\begin{multline} \label{eq:gen2}
b_m^{(i)}\sim\mbox{Bern}(\nu_m), \\
\nu_m\sim\mbox{Beta}(\alpha/M,1),~\gamma_0\sim\mbox{Ga}(c_0,1/d_0)
\end{multline}
where $\mbox{Ga}(\cdot)$ denotes the gamma distribution, and $\mbox{Bern}(\cdot)$ the Bernoulli distribution. Note that $\{\phi_m,\nu_m\}_{m=1,M}$ are shared across all intervals $i$, and it is in this manner we achieve joint clustering across all {time} intervals. 
The reasons for the choices of these various priors is discussed in Section \ref{sec:related}, when making connections to related models. For example, the choice $b_m^{(i)}\sim\mbox{Bern}(\nu_m)$ with $\nu_m\sim\mbox{Beta}(\alpha/M,1)$ is motivated by the connection to the Indian buffet process \cite{IBP} as $M\rightarrow\infty$.


We refer to this as a focused mixture model (FMM) because the $\nu_m$ defines the probability with which cluster $m$ is observed, and via the prior in (\ref{eq:gen2}) the model only ``focuses'' on a small number of clusters, those with large $\nu_m$. Further, as discussed below, the parameter $\phi_m$ controls the firing rate of neuron/cluster $m$, and that is also modeled. Concerning models to which we compare, when the $\pi_m^{(i)}$ are modeled via a Dirichlet process (DP) \cite{Antoniak74}, and the matrix of multi-channel data are modeled jointly, we refer to the model as matrix DP (MDP). If a DP is employed separately on each channel the results are simply termed DP. The hierarchical DP model in \cite{Bo2011} for $\pi_m^{(i)}$ the model is referred to as HDP.

\subsection{{Computations}}\label{sec:computations}

The posterior distribution of model parameters is approximated via Gibbs sampling. Most of the update equations for the model are relatively standard due to conjugacy of consecutive distributions in the hierarchical model; these ``standard'' updates are not repeated here (see \cite{Bo2011}). Perhaps the most important update equation is for $\phi_m$, as we found this to be a critical component of the success of our inference. To perform such sampling we utilize the following lemma.
\begin{lem}\label{lem:NBinference} Denote $s(n,j)$ as the Sterling numbers of the first kind \cite{johnson2005univariate} and $F(n,j) = (-1)^{n+j}s(n,j)/n!$ as their normalized and unsigned representations, with $F(0,0)=1$, $F(n,0) = 0$ if $n>0$, $F(n,j)=0$ if $j>n$ and
$F(n+1,j) =\frac{n }{n+1}F(n,j) + \frac{1}{n+1}F(n,j - 1)$
if $1\le j\le n$. Assuming $n\sim\emph{\mbox{NegBin}}(\phi,p)$ is a negative binomial distributed random variable, and it is augmented into a compound Poisson representation \cite{Anscombe1949} as  \beq n  = \sum_{l=1}^{\ell} u_{l},~ u_{l}\sim \emph{\mbox{Log}}(p),~ \ell\sim\emph{\mbox{Pois}}(-\phi\ln(1-p))\eeq where $\emph{\mbox{Log}}(p)$ is the logarithmic distribution \cite{Anscombe1949}  with probability generating function $G(z)=
{\ln(1-pz)}/{\ln(1-p)},~ |z|<{p^{-1}}$, then we have
\beq
\emph{\mbox{Pr}}(\ell= j | n,\phi) = R_{\phi}\left(n,j\right) =  {F(n,j) \phi^{j} }\bigg/{{\sum_{j'=1}^{n}F(n,j') \phi^{j'} }}\eeq for $j=0,1,\cdots,n$.

\end{lem}

The proof is provided in the Appendix.

{Let the total set of data measured during interval $i$ be represented $\bm{\mathcal{D}}_i=\{\Xmat_{ij}\}_{j=1}^{M_i}$, where $M_i$ is the total number of events during interval $i$.  Let $n_{im}^*$ represent the number of data samples in $\bm{\mathcal{D}}_i$ that are apportioned to cluster $m\in\{1,\dots,M\}=\mathcal{S}$, with $M_i$$=\sum_{m=1}^M n_{im}^*$.}
{To sample} $\phi_m$, since p($\phi_m|{\pv},n_{\cdot m}^\star)\propto$ $ \prod_{i: b_m^{(i)}=1}$ $\mbox{NegBin}(n^{*}_{im};\phi_m ,p_i)\mbox{Ga}( \phi_m;\gamma_0,1)$ {(see Appendix}  \ref{sec:related} { for details)}, using Lemma \ref{lem:NBinference}, we can first sample a latent count variable $\ell_{im}$ for each $n^{*}_{im}$ as
\beq
\mbox{Pr}(\ell_{im} = l|n^{*}_{im},\phi_m) = R_{\phi_m}(n^*_{im},l),~~l=0,\cdots, n^*_{im}.
\eeq
Since $\ell_{im}\sim \mbox{Pois}(-\phi_m\ln(1-p_i))$, using the conjugacy between the gamma and Poisson distributions, we have
\beqs
& \phi_m|\{\ell_{im},b_m^{(i)},p_i\}  \sim \nonumber\\& \mbox{Ga} \left( \gamma_0 +  \sum_{i: b_m^{(i)}=1}  \ell_{im}, \frac{1}{1 - \sum_{i: b_m^{(i)}=1} \ln( 1 - p_i)}\right).
\eeqs
Notice that marginalizing out $\phi_m$ in $\ell_{im}\sim \mbox{Pois}(-\phi_m\ln(1-p_i))$ results in $\ell_{im}\sim \mbox{NegBin}(\gamma_0,\frac{-\ln(1-p_i)}{1-\ln(1-p_i)})$, therefore, we can use the same data augmentation technique by sampling a latent count $\tilde{\ell}_{im}$ for each $\ell_{im}$ and  then 
sampling $\gamma_0$ using the gamma Poisson conjugacy as
\beqs
&\mbox{Pr}(\tilde{\ell}_{im} = l|\ell_{im},\gamma_0) = R_{\gamma_0}(\ell_{im},l),~~l=0,\cdots, \ell_{im}\\
&\gamma_0|\{\tilde{\ell}_{im},b_m^{(i)},p_i\}  \sim \nonumber\\&\mbox{Ga} \left( c_0 +  \sum_{i: b_m^{(i)}=1}  \tilde{\ell}_{im}, \frac{1}{d_0 - \sum_{i: b_m^{(i)}=1} \ln\big( 1 - \frac{-\ln(1-p_i)}{1-\ln(1-p_i)}\big)}\right)\nonumber.
\eeqs

Another important parameter is $b_m^{(i)}$.  Since $b_m^{(i)}$ can only be zero if $n^*_{im}=0$ and when $n^*_{im}=0$, $\mbox{Pr}( b_m^{(i)}=1|-)\propto \mbox{NegBin}(0;\phi_m ,p_i)\pi_m$ and $\mbox{Pr}( b_m^{(i)}=0|-)\propto (1-\pi_m)$,  we have
\beqs
& b_m^{(i)}| \pi_m, n^*_{im},\phi_m,p_i \sim \nonumber\\ &\mbox{Bernoulli}\left(\delta(n^*_{im}=0) \frac{\pi_m(1-p_i)^{\phi_m}}{\pi_m(1-p_i)^{\phi_m} + (1-\pi_m)} + \delta(n^*_{im}>0)\right).\nonumber
\eeqs
A large $p_i$ thus indicates a large variance-to-mean ratio on $n_{im}^*$ and $M_i$. Note that when $b_m^{(i)}=0$, the observed zero count $n_{im}^*=0$ is no longer explained by $n_{im}^*\sim \mbox{NegBin}(r_m,p_i)$, this satisfies the intuition that the underlying beta-Bernoulli process is governing whether a cluster would be used or not, and once it is activated, it is $r_m$ and $p_i$ that control how much it would be used.

\subsection{Data Acquisition and Pre-processing} 
\label{sub:data_acquisition_and_pre_processing}

In this work we use two datasets, the popular ``hc-1'' dataset\footnote{available from http://crcns.org/data-sets/hc/hc-1} and a new dataset based upon experiments we have performed with freely moving rats (institutional review board approvals were obtained). These data will be made available to the research community.  Six animals were used in this study. Each animal was trained, under food restriction (15 g/animal/day, standard hard chow), on a simple lever-press-and-hold task until performance stabilized and then taken in for surgery.  Each animal was implanted with four different silicon microelectrodes  (NeuroNexus Technologies; Ann Arbor, MI; custom design) in the forelimb region of the primary or supplementary motor cortex. Each electrode contains up to 16 independent recording sites, with variations in device footprint and recording site position ($e.g.$, Figure \ref{fig:device}).  Electrophysiological data were measured during one-hour periods on eight consecutive days, starting on the day after implant (data were collected for additional days, but the signal quality degraded after 8 days, as discussed below).  The recordings were conducted in a high walled observation chamber under freely-behaving conditions.  Note that nearby sensors are close enough to record the signal of a single or small group of neurons, termed a single-unit event. However, in the device in Figure \ref{fig:device}, all eight sensors in a line are too far separated to simultaneously record a single-unit event on all eight.

The data were bandpass filtered (0.3-3 kHz), and
then all signals 3.5 times the standard deviation of the background
signal were deemed detections. The peak of the detection was placed
in the center of a 1.3 msec window, which corresponds to $T=40$
samples at the recording rate. The signal
$\Xmat_{ij}\in\mathbb{R}^{T\times N}$ corresponds to the data
measured simultaneously across all $N$ channels within this window.
Here $N=8$, with a concentration on the data measured from the 8
channels of the zoomed-in Figure \ref{fig:device}.



\subsection{Evaluation Criteria} \label{sec:eval}

We use several different criteria to evaluate the performance of the competing methodologies. 
{Let  $F_p$ and $F_n$ denote the total number of false positives and negatives for a given neuron, respectively, and let $\#_w$ denote the total number of detected waveforms.}
{We define:}
\begin{align}
	\mbox{Accuracy} = \left\{1-\frac{F_p+F_n}{\#_w}\right\} \times 100\%.
\end{align}
For synthetic missing data, as in Section \ref{sec:missing}, we compute the relative recovery error (RRE):
\begin{align}
	\mbox{RRE} = \frac{\norm{\mb{X}-\mhb{X}}}{\norm{\mb{X}}} \times 100\%, 
\end{align}
where $\mb{X}$ is the true waveform, $\mhb{X}$ is the estimated waveform, and $\norm{\cdot}$ indicates the $L_2$ or Frobenius norm depending on context. 
When adding noise, we compute the signal-to-noise ratio (SNR) as in \cite{Suner2005}:
\begin{align}
	\text{SNR} = \frac{A}{2 SD_{noise}},	
\end{align}
where $A$ denotes the peak-to-peak voltage difference of the mean waveform and $SD_{noise}$ is the standard deviation of the noise.  The noise level is estimated by mean absolute deviation.

To simulate a lower SNR in the sparse spiking experiments, we took background signals from the  dataset where no spiking occurred and scale them by $\alpha$ and add them to our detected spikes; this gives a total noise variance of $\sigma^2 (1+\alpha^2)$, and we set the SNR to 2.5 and 1.5 for these experiments.

\section{Results\label{sec:results}}

For these experiments we used a truncation level of
$K=40$ dictionary elements, and the number of mixture components was
truncated to $M=20$ (these truncation levels are upper bounds, and within the analysis a subset of the possible dictionary elements and mixture components are utilized).  In dictionary learning, the gamma priors for
$\{\eta_t\}$ and $\alpha_0$ were set as
$\mbox{Ga}(10^{-6},10^{-6})$. In the context of the {focused mixture model}, we set $a_0=b_0=1$, $c_0=0.1$ and
$d_0=0.1$. Prior $\mbox{Ga}(10^{-6},10^{-6})$ was placed on
parameter $\alpha$ related to the {Indian Buffet Process (see Appendix}  \ref{sec:related} {for details)}. None of these parameters have
been tuned, and many related settings yield similar results. In all
examples we ran 6,000 Gibbs samples, with the first 3,000 discarded
as burn-in (however, typically high-quality results are inferred with far fewer samples, offering the potential for computational acceleration).

\subsection{Real data with partial ground truth} 
\label{sec:truth}
We
first consider publicly available dataset
hc-1. These data consist of both
extracellular recordings and an intracellular recording from a
nearby neuron in the hippocampus of an anesthetized rat
\cite{Henze2000}.  Intracellular recordings give clean signals on a
spike train from a specific neuron, providing accurate spike times for
that neuron.  Thus, if we detect a spike in a nearby extracellular
recording within a close time period ($<0.5$ms) to an intracellular
spike, we assume that the spike detected in the extracellular
recording corresponds to the known neuron's spikes.

\begin{figure}[h!]
  \centering
    \includegraphics[width=1.0\linewidth]{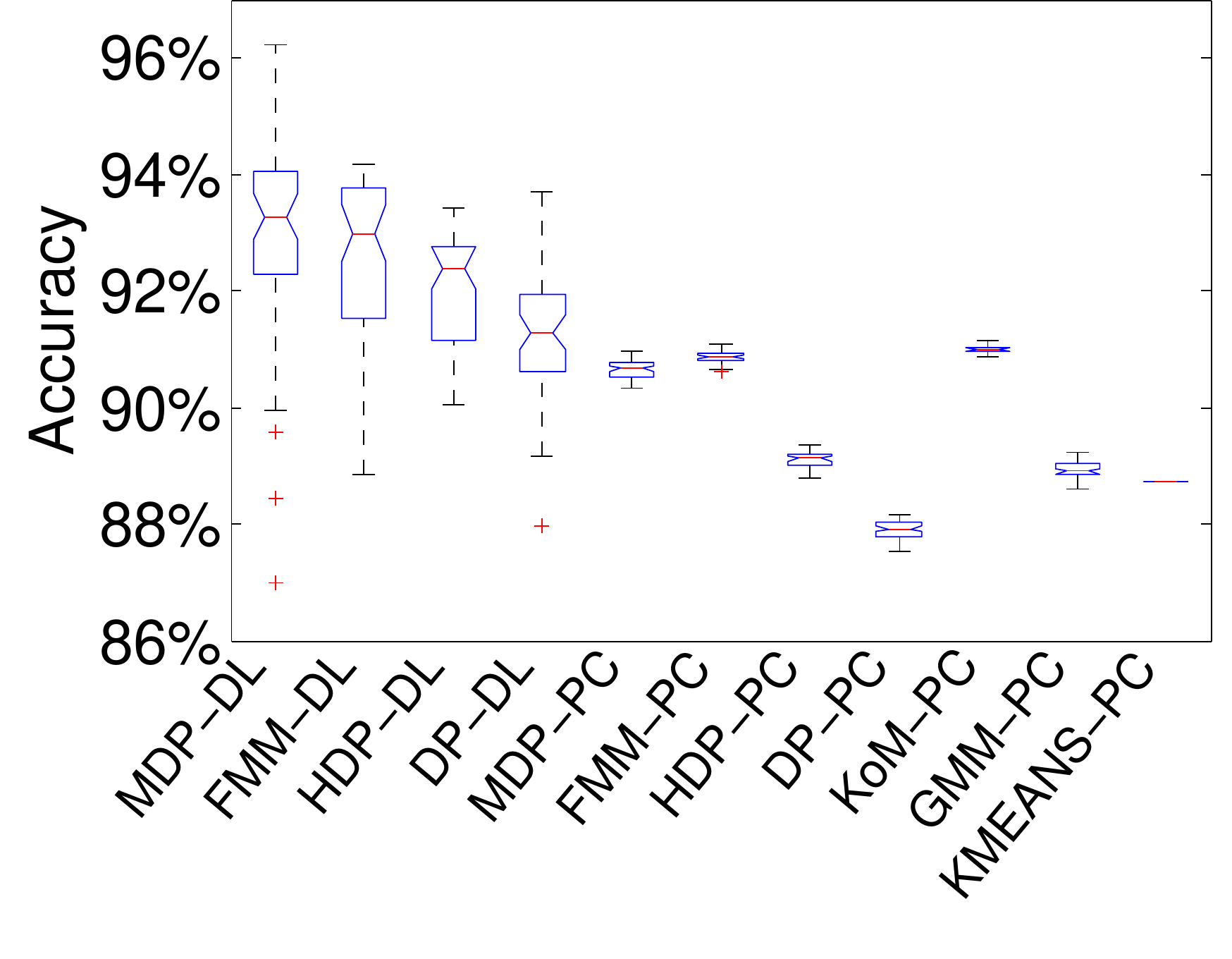}
    \caption{{{ Accuracy of the various methods on d533101 data \cite{Henze2000}. All abbreviations are explained in the main text (Section \ref{sec:truth}).  Note that dictionary learning dominates performance over principal components.  Moreover, modeling multiple channels (as in MDP and FMM) dominates performance over operating on each channel separately.   
}}} \vspace{-10pt}
\label{fig:Accuracy_hc_1}
\end{figure}

We considered
the widely used data d533101 and the same
preprocessing from \cite{Calabrese2010}.
  These data consist of 4-channel extracellular recordings and 1-channel
  intracellular recording.  We used 2491 detected spikes and 786 of those
  spikes came from the known neuron. Accuracy of cluster results based on multiple methods are shown in Figure \ref{fig:Accuracy_hc_1}. 
We consider several different clustering schemes and two different strategies for learning low-dimensional representations of the data.  Specifically, we learn low-dimensional representations using either: dictionary learning (DL) or the first two principal components (PCs) of the matrix consisting of the concatenated waveforms.  For the multichannel data, we stack each waveform matrix to yield a vector, and concatenate stacked waveforms to obtain the data matrix upon which PCA is run.   Given this representation, we consider several different clustering strategies: (i) Matrix Dirichlet Process (MDP), which implements a DP on the $\Xmat_{ij}$ \emph{matrices}, as opposed to previous DP approaches on vectors \cite{Wood2009, Bo2011}, (ii) focused mixture model (as described above), (iii) Hierarchical DP (HDP) \cite{Bo2011}, (iv) independent DP (the HDP and independent DP are from \cite{Bo2011}), (v) Mixture of Kalman filters (MoK) \cite{Calabrese2010}, (vi) Gaussian mixture models (GMM) \cite{bishop2006}, and (vii) K-means (KMEANS) \cite{Lewicki}.  Although we do not consider all pairwise comparisons, we do consider many options.  Note that all of the DL approaches are novel.  
It should be clear from Figure \ref{fig:Accuracy_hc_1} that dictionary learning enhances performance over principal components for each clustering approach. Specifically, all DL based methods outperform all PC based methods.
Moreover, sharing information across channels, as in MDP and FMM (both novel methodologies), seems to further improve performance.   The ordering of the algorithms is essentially unchanged upon using a different number of mixture components or a different number of principal components.  

\begin{figure}[h!]
  \centering
    \includegraphics[width=1.0\linewidth]{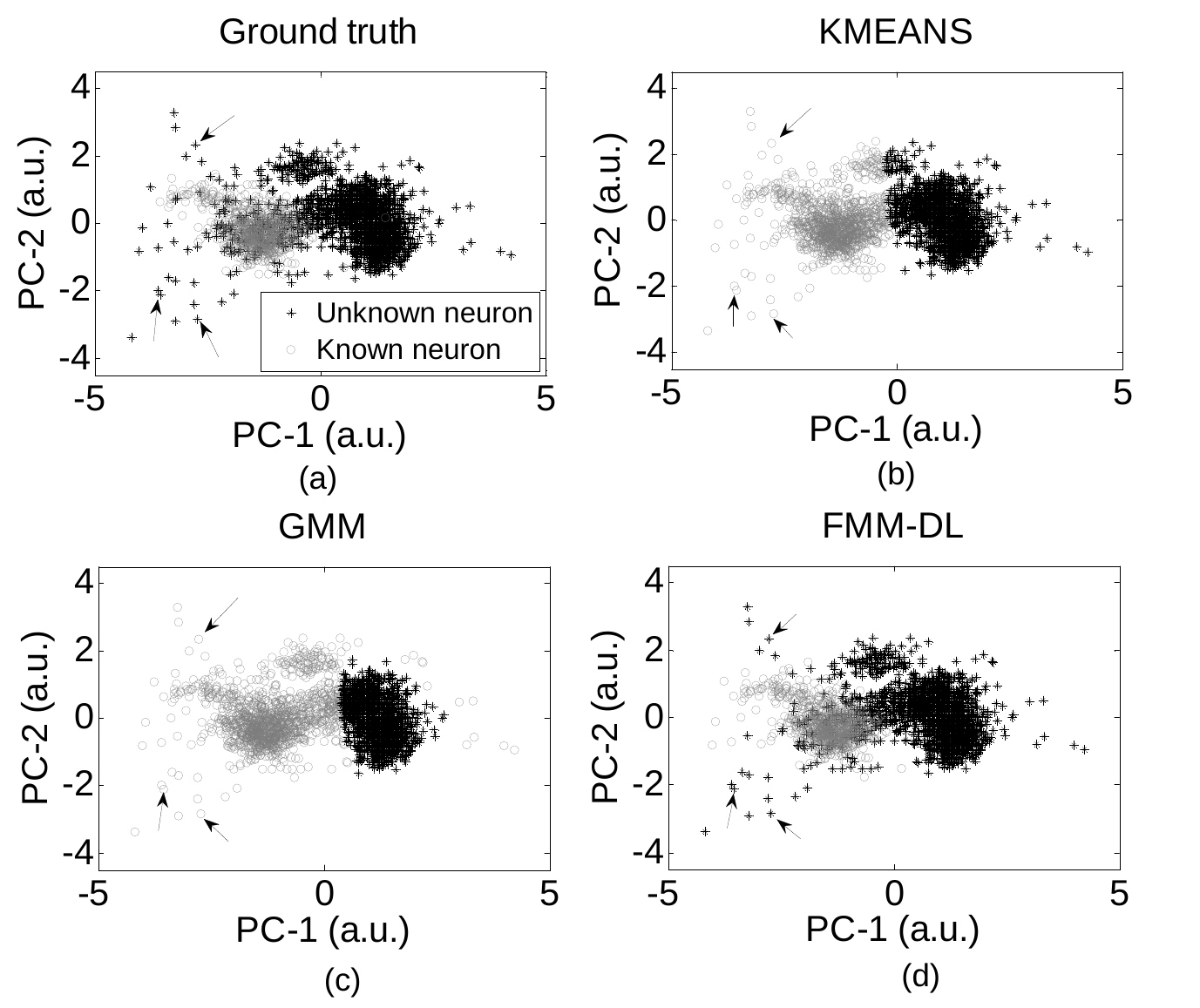}
    \caption{{{ Clustering results shown in the 2 PC space of the various methods on d533101 data \cite{Henze2000}. All abbreviations are explained in the main text (Section \ref{sec:truth}). ``Known neuron" denotes waveforms associated with the neuron from the cell with the intracellular recording, and ``Unknown neuron'' refers to all other detected waveforms.   Note that all methods are shown in the first two PCs for visualization, but that the FMM-DL shown in (d) is jointly learning the feature space and clustering.
}}} \vspace{-10pt}
\label{fig:Dictionary_hc_1}
\end{figure}

In Figure \ref{fig:Dictionary_hc_1}, we visualize the waveforms in the first 2 principle components for comparison.  In Figure \ref{fig:Dictionary_hc_1}a, we show ground truth to compare to the results we get by clustering from the K-means algorithm shown in Figure \ref{fig:Dictionary_hc_1}b and the clustering from the GMM shown in  Figure \ref{fig:Dictionary_hc_1}c.  We observe that both K-means and GMM work well, but due to the constrained feature space they incorrectly classify some spikes (marked by arrows). However, the proposed model, shown in Figure  \ref{fig:Dictionary_hc_1}(d), which incorporates dictionary learning with spike sorting, infers an appropriate feature space (not shown) and more effectively clusters the neurons.


Note that in Figure \ref{fig:Accuracy_hc_1} and  \ref{fig:Dictionary_hc_1}, in the context of PCA features, we considered the two principal components (similar results were obtained with the three principal components); when we considered the 20 principal components, for comparison, the results deteriorated, presumably because the higher-order components correspond to noise. An advantage of the proposed approach is that we model the noise explicitly, via the residual $\Emat_{ij}$ in (\ref{eq:basic}); with PCA the signal and noise are not explicitly distinguished. 


\subsection{Longitudinal analysis of electrophysiological data\label{sec:forensics}}

\begin{figure}[!htbp]
\centering

\subfigure[]{
   \includegraphics[width=0.7\linewidth] {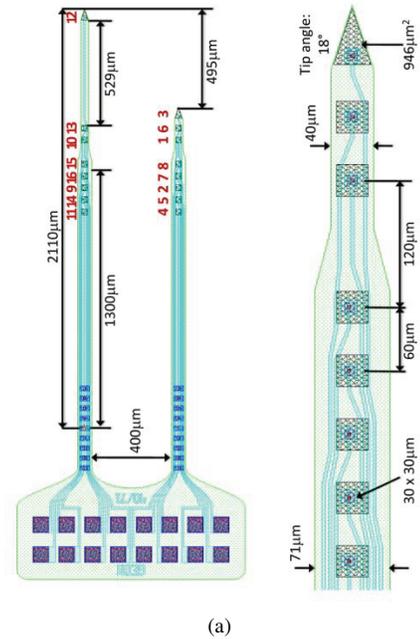}
   \label{fig:device}

 }
 \subfigure[]{
   \includegraphics[width=0.7\linewidth] {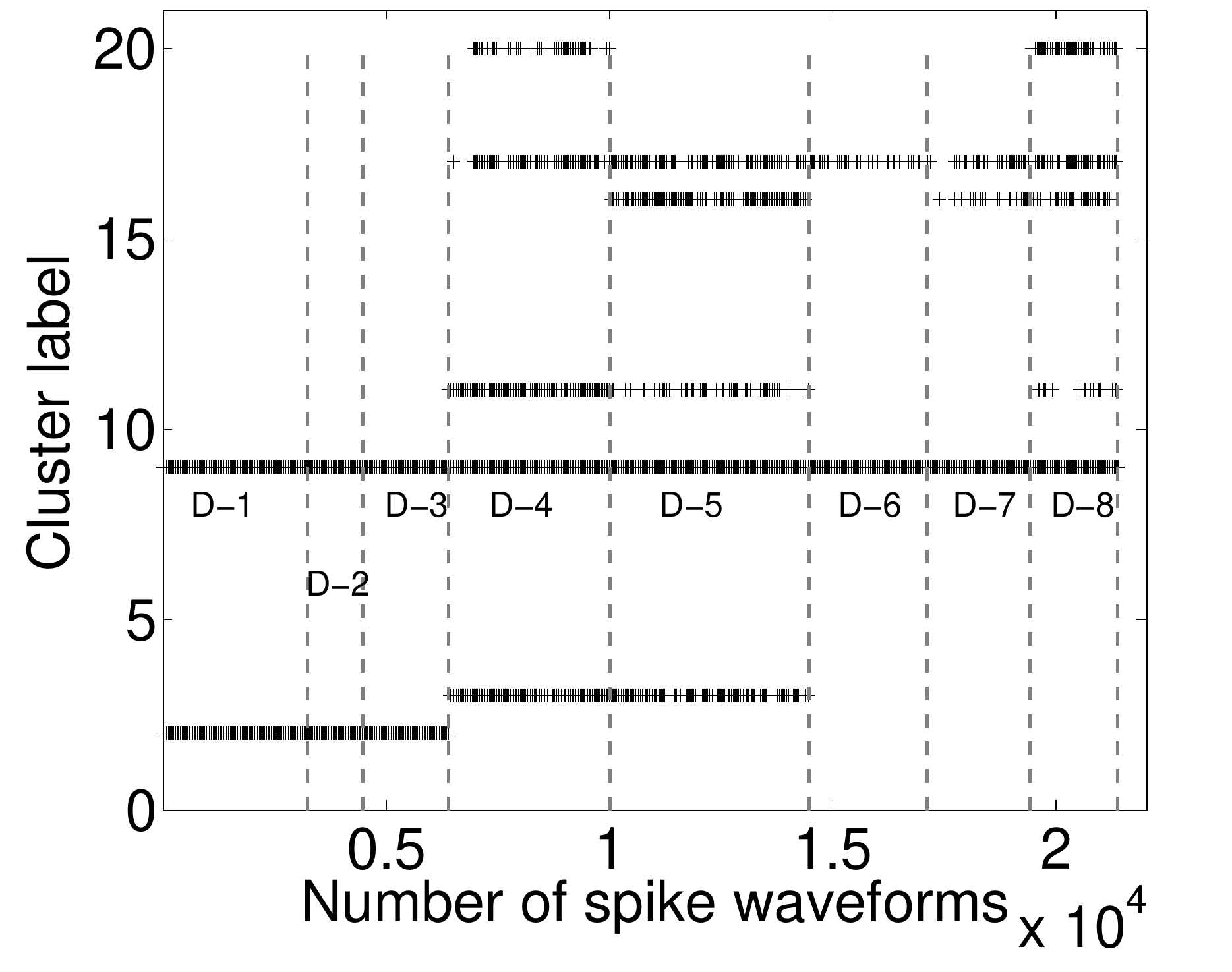}
   \label{fig:glob_clustering}
 }
 \subfigure[]{
   \includegraphics[width=0.7\linewidth] {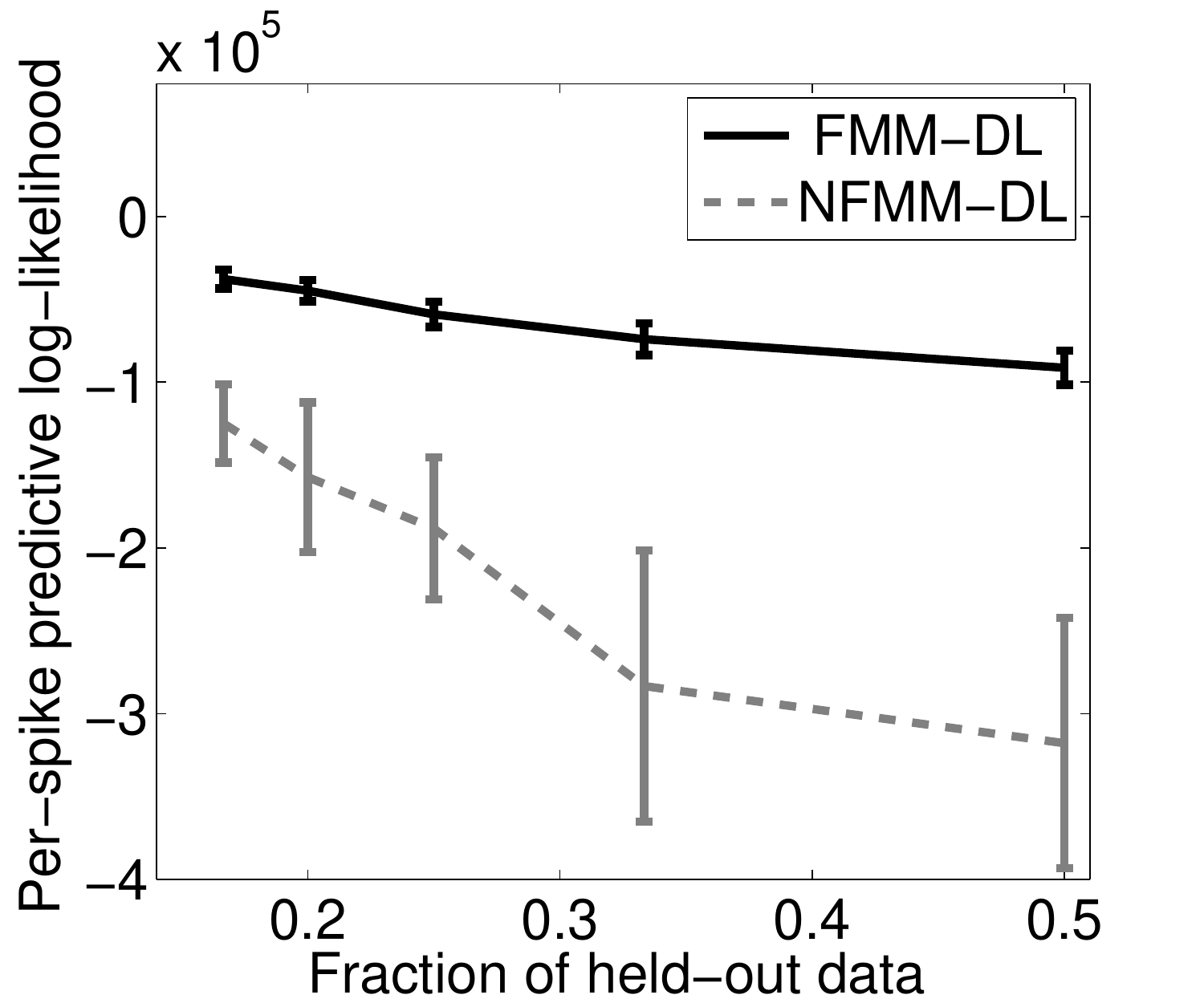}
   \label{fig:likelihood}
 }
  \caption{{
{Longitudinal data analysis of the rat motor cortex data. }
(a) Schematic of the neural
recording array that was placed in the rat motor cortex. The red
numbers identify the sensors, and a zoom-in of the bottom-eight
sensors is shown. The sensors are ordered by the order of the
read-out pads, at left. The presented data are for sensors numbered 1 to 8, corresponding to the zoomed-in region. (b) From the maximum-likelihood collection
sample, the apportionment of data among mixture components
(clusters). Results are shown for 45 sec recording periods, on each
of 8 days. For example, D-4 reflects data on day 4. Note that while the truncation level is such that there are 20 candidate clusters (vertical axis in (b)), only an inferred subset of clusters are actually used on any given day. (c) Predictive likelihood of held-out data. The
horizontal axis represents the fraction of data held out during training. 
{FMM-DL dominates NFMM-DL on these data. }}} \label{fig:long} \vspace{-20pt}
\end{figure}

{Figure }\ref{fig:glob_clustering}{(a) shows the recording probe used for the analysis of the rat motor cortex data.}  Figure \ref{fig:glob_clustering} {shows} assignments of data to each of the possible clusters, for data measured across the 8 days, as computed by the proposed model (for example, for the first three days, two clusters were inferred). Results are shown for the maximum-likelihood collection sample. As a comparison to {FMM-DL,} 
we also considered the {non-focused mixture model (NFMM-DL)} methodology discussed in Section \ref{sec:related}, with the $\bv^{(i)}$ set to all ones (in both cases we employ the same form of dictionary learning, as in Section \ref{sec:dict}). From Figure \ref{fig:likelihood}, it is observed that on held-out data the FMM{-DL} yields improved results relative to the {NFMM-DL}.

In fact, the proposed model was developed specifically to address the problem of multi-day {longitudinal} analysis of electrophysiological data, as a consequence of observed limitations of HDP (which are only partially illuminated by Figure \ref{fig:likelihood}). Specifically, while the focused nature of the FMM{-DL} allows learning of specialized clusters that occur over limited days, the ``non-focused'' HDP{-DL} tends to merge similar but distinct clusters. This yields HDP results that are characterized by fewer total clusters, and by cluster characteristics that are less revealing of detailed neural processes. Patterns of observed neural activity may shift over a period of days due to many reasons, including cell death, tissue encapsulation, or device movement; this shift necessitates the FMM{-DL}'s ability to focus on subtle but important differences in the data properties over days. This ability to infer subtly different clusters is related to the focused topic model's ability  \cite{compound} to discern distinct topics that differ in subtle ways. The study of large quantities of data (8 days) makes the ability to distinguish subtle differences in clusters more challenging (the DP-DL-based model works well when observing data from one recording session, like in Figure \ref{fig:Accuracy_hc_1}, but the analysis of multiple days of data is challenging for HDP).

\begin{figure}[!htbp]
\centering

\subfigure[]{
   \includegraphics[width=0.7\linewidth] {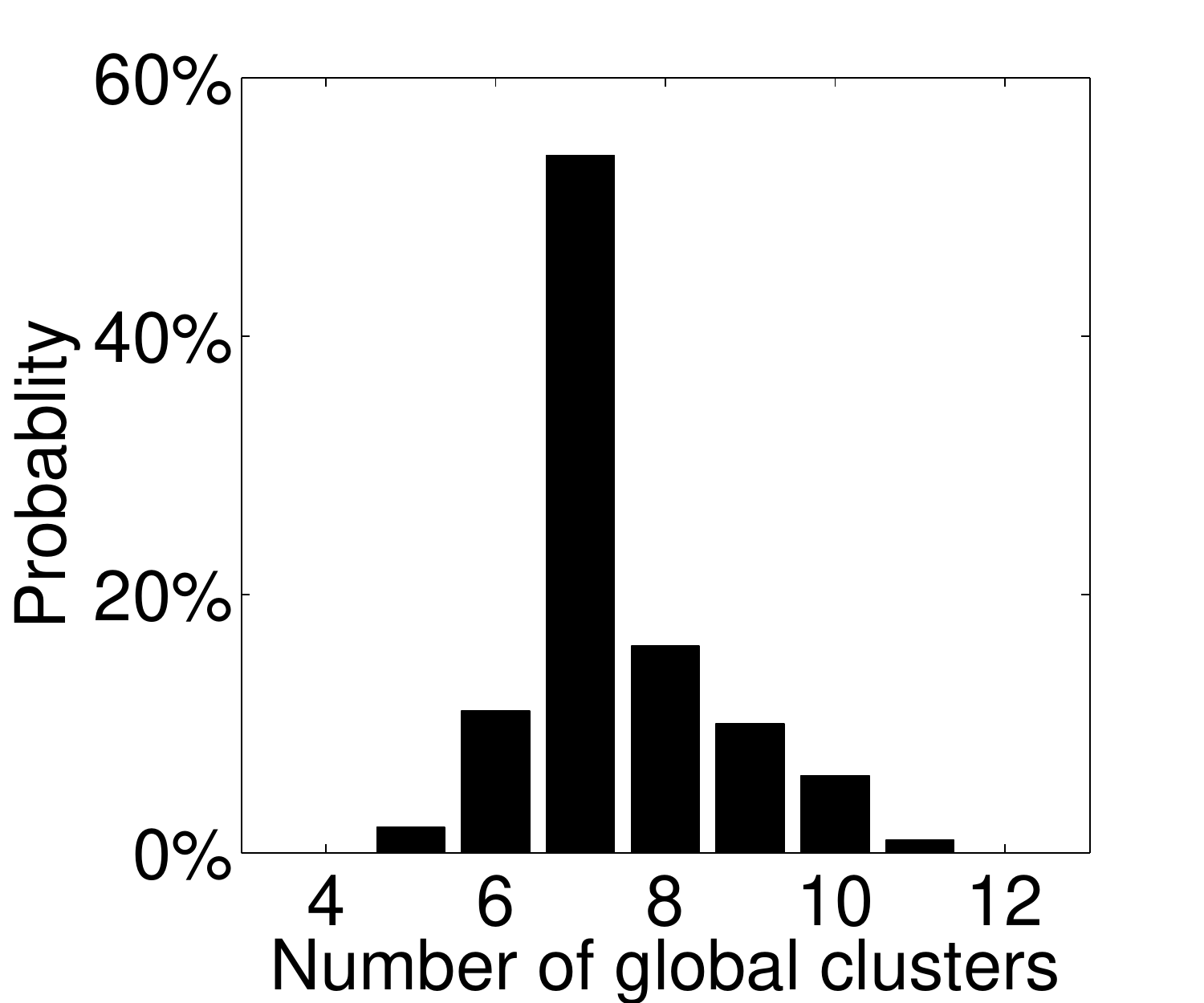}
   \label{fig:post_clusters}
 }
 \subfigure[]{
   \includegraphics[width=0.7\linewidth] {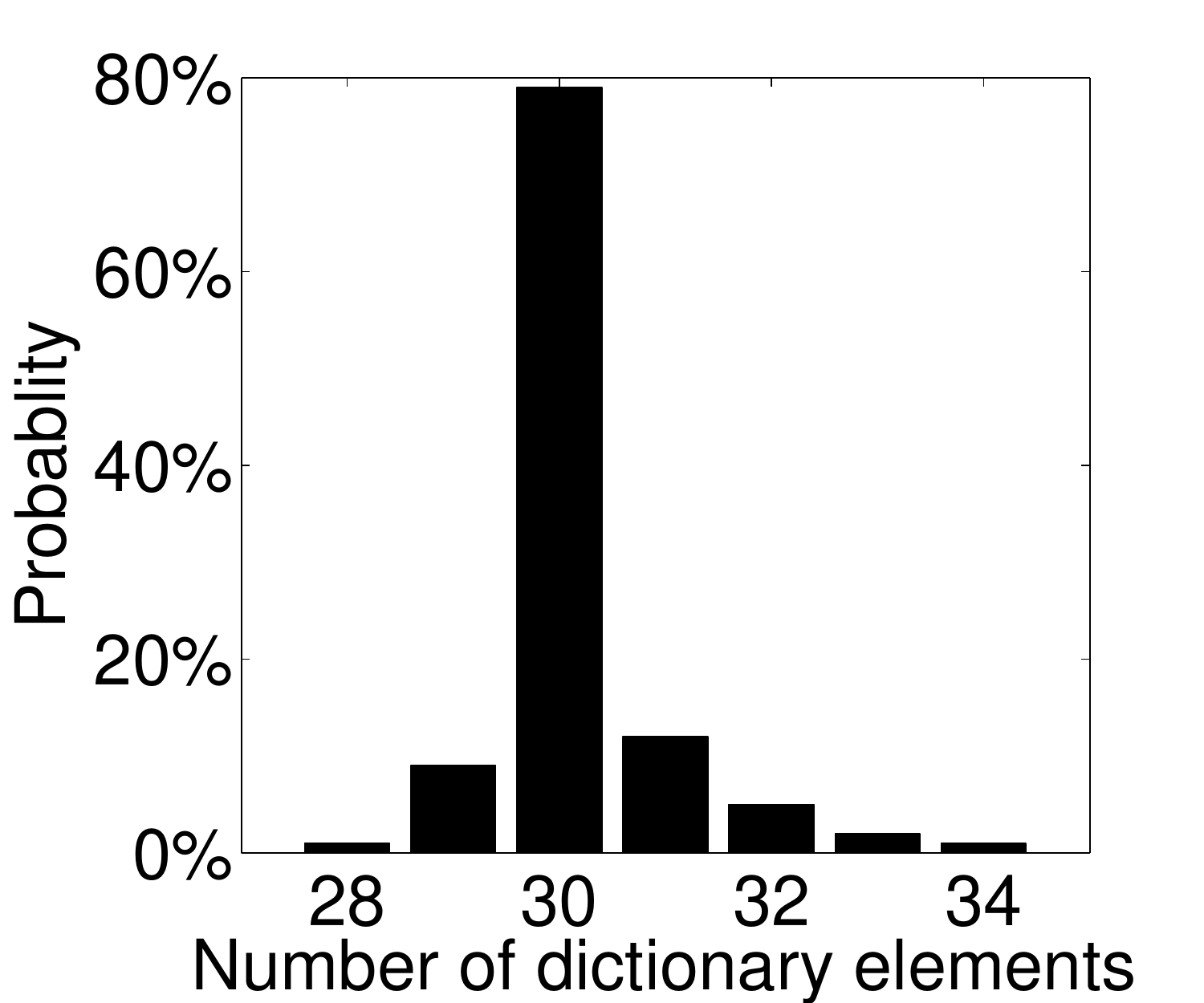}
   \label{fig:post_dict}
 }\hspace{-5mm}
 \subfigure[]{
   \includegraphics[width=1.0\linewidth] {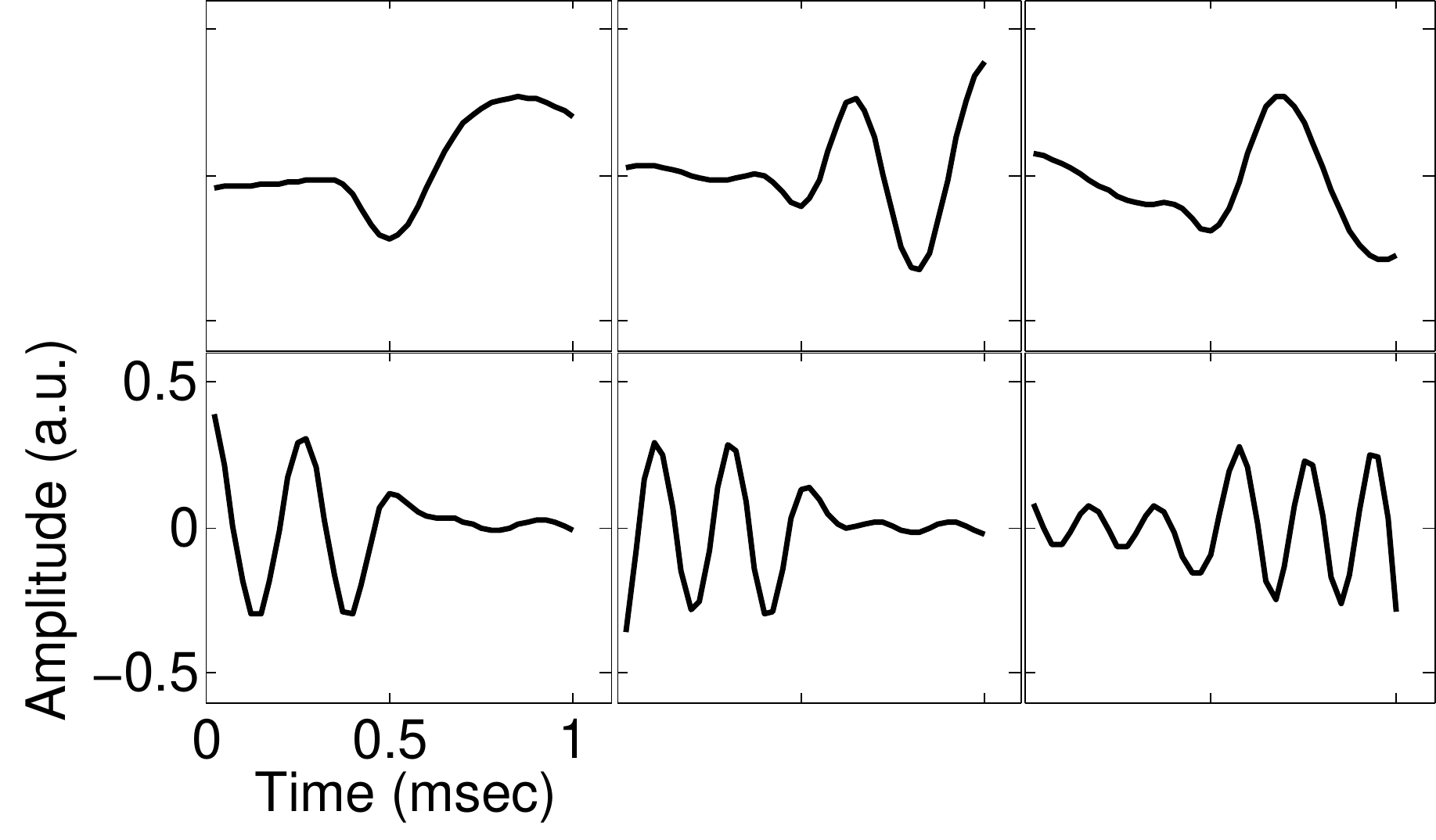}
   \label{fig:dict_examples}
 }
  \caption{{
{Posteriors and dictionaries from rat motor cortex data (the same data as in Figure} \ref{fig:long}).(a) Approximate posterior distribution on the number of global clusters (mixture components). (b) Approximate posterior distribution of the
number of dictionary elements. (c) Examples of inferred
dictionary elements{; amplitudes of dictionary elements are unit less.} }} \vspace{-20pt}
\end{figure}

Note from Figure \ref{fig:glob_clustering} that the number of
detected signals is different for different recording days, despite
the fact that the recording period reflective of these data (45
secs) is the same for all days. This highlights the need to allow
modeling of different {firing} rates, as in our model but not
emphasized in these results.

Among the parameters inferred by the model are approximate posterior
distributions on the number of clusters across all days, and on the
required number of dictionary elements. These approximate posteriors
are shown in Figures \ref{fig:post_clusters}{ and }\ref{fig:post_dict},
and Figure \ref{fig:dict_examples} {shows } example dictionary
elements. Although not shown for brevity, the $\{p_i\}$ had posterior means in excess of 0.9.

\begin{figure}[h!]
  \centering
    \includegraphics[width=1.0\linewidth]{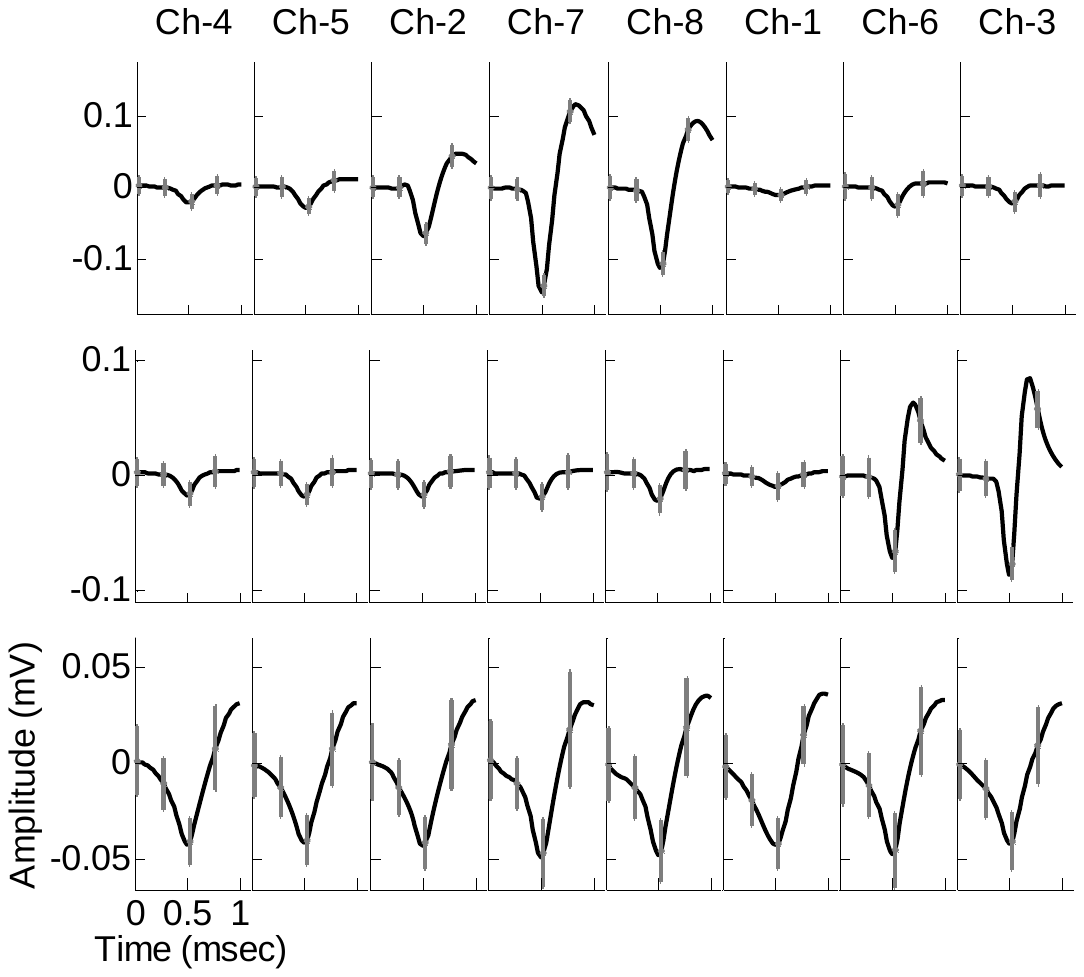}
\caption{{{Example
clusters inferred for data on the bottom 8 channels of Fig.
\ref{fig:device}. (a)-(b) Example of single-unit events. (c) Example
of a cluster \emph{not} attributed to a single-unit-event. The 8
signals are ordered from left to right consistent with the numbering
of the 8 channels at the bottom of Figure \ref{fig:device}. The black curves represent the mean, and the error bars are one standard deviation.}}}
\label{fig:units}
\end{figure}

To better represent insight that is garnered from the model,  Figure \ref{fig:units} {depicts } the inferred properties of three of the clusters, from Day 4 (D-4 in Figure \ref{fig:glob_clustering}). Shown are the \emph{mean} signal for the 8 channels in the respective cluster (for the 8 channels at the bottom of Figure \ref{fig:device}), and the error bars represent one standard deviation, as defined by the estimated posterior. Note that the cluster in {the top row of} Figure \ref{fig:units} corresponds to a localized single-unit event, presumably from a neuron (or a coordinated small group of neurons) near the sensors associated with channels 7 and 8. The cluster in {the middle row of } Figure \ref{fig:units} similarly corresponds to a single-unit event situated near the sensors associated with channels 3 and 6. Note the proximity of sensors 7 and 8, and sensors 3 and 6, from Figure \ref{fig:device}. The HDP model uncovered the cluster in {the top row of } Figure \ref{fig:units}, but not that in {the middle row of } Figure \ref{fig:units} {(not shown)}.

Note {the bottom row of} Figure \ref{fig:units}, in which the mean signal across all 8 channels is approximately the same (HDP also found related clusters of this type). This cluster is deemed to \emph{not} be associated with a single-unit event, as the sensors are too physically distant across the array for the signal to be observed simultaneously on all sensors from a single neuron. This class of signals is deemed associated with an artifact or some global phenomena, (possibly) due to movement of the device within the brain, and/or because of charges that build up in the device and manifest signals with animal motion (by examining separate video recordings, such electrophysiological data occurred when the animal constituted significant and abrupt movement, such as heading hitting the sides of the cage, or during grooming). Note that in {the top two rows of} Figure \ref{fig:units} the error bars are relatively tight with respect to the strong signals in the set of eight, while the error bars in Figure \ref{fig:units}(c) are more pronounced (the mean curves look {smooth}
, but this is based upon averaging thousands of signals).

In addition to recording the electrophysiological data, video was recorded of the rat throughout{ the experiment}. Robust PCA \cite{Wright09} was used to quantify the change in the video from frame-to-frame, with high change associated with large motion by the animal (this automation is {useful} because one hour of data are collected on each day; direct human viewing is tedious and unnecessary). On Day 4, the model infers that in periods of high animal activity, 20\% to 40\% of the detected signals are due to single-unit events (depending on which portion of data are considered); during periods of relative rest 40\% to 70\% of detected signals are due to single-unit events. This suggests that animal motion causes signal artifacts, as discussed in Section \ref{sec:intro}

In these studies the total fraction of single-unit events, even when at rest, diminishes with increasing number of days from sensor implant; this may be reflective of changes in the system due to the glial immune response of the brain \cite{Biran,Szarowski03}. The discerning ability of the proposed FMM{-DL} to distinguish subtly different signals, and analysis of data over multiple days, has played an important role in this analysis. Further, {longitudinal} analyses like that in Figure \ref{fig:units} were the principal reason for modeling the data on all $N=8$ channels jointly (the ability to distinguish single-unit events from anomalies is predicated on this multi-channel analysis).

\subsection{Handling missing data} \label{sec:missing}

The quantity of data acquired by a neural recording system is enormous, and therefore in many systems one first performs spike detection (for example, based on a threshold), and then a signal is extracted about each detection (a temporal window is placed around the peak of a given detection). This step is often imperfect, and significant portions of many of the spikes may be missing due to the windowed signal extraction (and the missing data are not retainable, as the original data are discarded). Conventional feature-extraction methods typically cannot be applied to such temporally clipped signals.

Returning to (\ref{eq:basic}), this implies that some columns of the data $\Xmat_{ij}$ may have missing entries. Conditioned on $\Dmat$, $\Lambdamat$, $\Smat_{ij}$, and $(\eta_1,\dots,\eta_T)$, we have $\Xmat_{ij}\sim\mathcal{N}(\Dmat\Lambdamat\Smat_{ij},\mbox{diag}(\eta_1^{-1},\dots,\eta_T^{-1})$. The missing entries of $\Xmat_{ij}$ may be treated as random variables, and they are integrated out analytically within the Gaussian likelihood function. Therefore, for the case of missing data in $\Xmat_{ij}$, we simply evaluate (\ref{eq:basic}) at the points of $\Xmat_{ij}$ for which data are observed. The columns of the dictionary $\Dmat$ of course have support over the entire signal, and therefore given the inferred $\Smat_{ij}$ (in the presence of missing data), one may impute the missing components of $\Xmat_{ij}$ via $\Dmat\Lambdamat\Smat_{ij}$. As long as, across all $\Xmat_{ij}$, the same part of the signal is not clipped away (lost) for all observed spikes, by jointly processing all of the {retained} data (all spikes) we may infer $\Dmat$, and hence infer missing data.

In practice we are less interested in observing the imputed missing parts of $\Xmat_{ij}$ than we are in simply clustering the data, in the presence of missing data. By evaluating $\Xmat_{ij}\sim\mathcal{N}(\Dmat\Lambdamat\Smat_{ij},\mbox{diag}(\eta_1^{-1},\dots,\eta_T^{-1}))$ only at points for which data are observed, and via the mixture model in (\ref{eq:mixture}), we directly infer the desired clustering, in the presence of missing data (even if we are not explicitly interested in subsequently examining the imputed values of the missing data).

\begin{figure}[!htbp]
\centering
\subfigure[]{
   \includegraphics[width=0.7\linewidth] {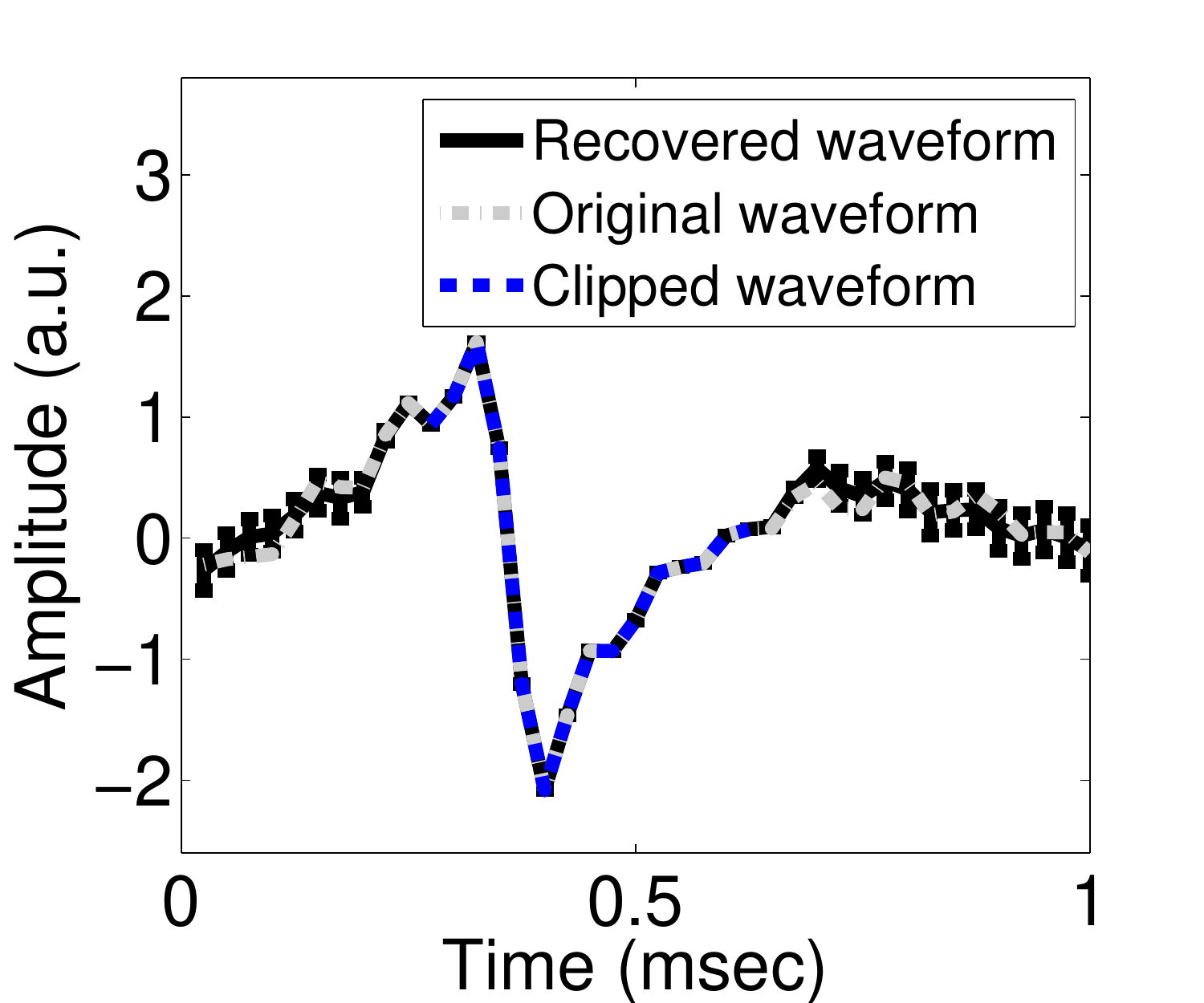}
   \label{fig:Recovery_waveform}
 } \subfigure[]{
   \includegraphics[width=0.7\linewidth] {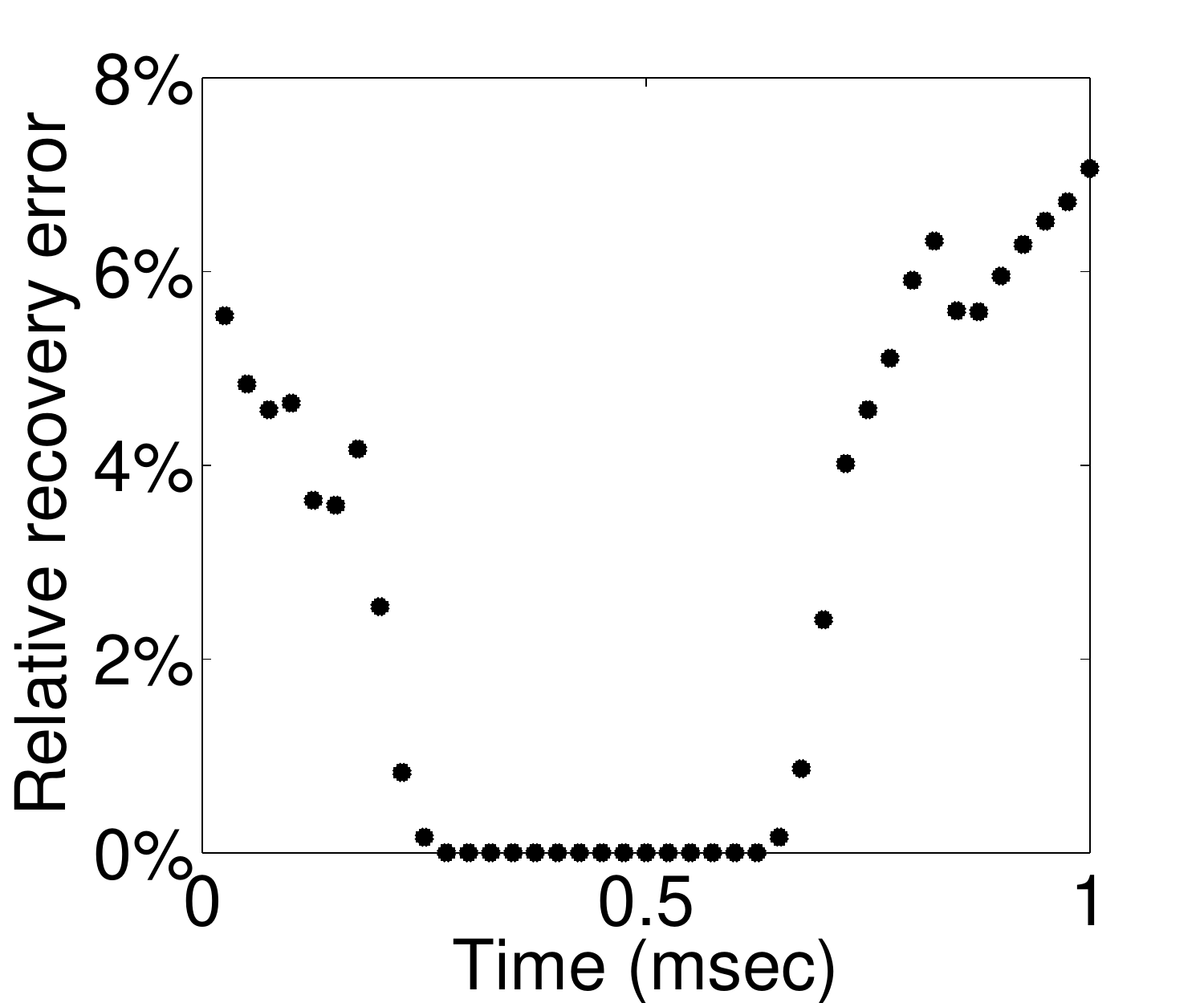}
   \label{fig:Relative_Error}
 }
  \caption{ { Our generative model elegantly addresses missing data.
(a) Example of a clipped waveform from the publicly available data (blue), original waveform (gray) and
  recovery waveform (black); the error bars reflect one standard deviation from the posterior distribution on the underlying signal. (b) Relative errors (with respect to the mean estimated signal). Note that we only show part of the waveform for visualization purposes.
   }} \vspace{-20pt} \label{fig:missing}
\end{figure}

To examine the ability of the model to perform clustering in the
presence of missing data, we reconsider the publicly available data
from Section \ref{sec:truth}. For the first 10\% of the spike
signals (300 spike waveforms), we impose that a fraction of
the beginning and end of the spike is absent. The original signals
are of length $T=40$ samples. As a demonstration, for the ``clipped'' signals, the first 10 and the last 16 samples of the
signals are missing. A clipped waveform example is shown in Figure \ref{fig:Recovery_waveform}; we compare the mean estimation of the signal, and the error bars reflect one standard deviation from the full posterior on the signal.
In the context of the analysis, we processed all of the data as before, but now with these ``damaged''/clipped signals. We observed that 94.11\% of the non-damaged signals were clustered properly (for the one neuron for which we had truth), and 92.33\% of the damaged signals were sorted properly. The recovered signal in Figure \ref{fig:Recovery_waveform} is typical, and is meant to give a sense of the accuracy of the recovered missing signal. The ability of the model to perform spike sorting in the presence of substantial missing data is a key attribute of the dictionary-learning-based framework. 


\subsection{Model tuning} \label{sec:tuning}

As constituted in Section \ref{sec:models}, the model is essentially parameter free. All of the hyperparameters are set in a relatively diffuse manner (see the discussion at the beginning of Section \ref{sec:results}), and the model infers the number of clusters and their composition with no parameter tuning required. Thus, our code runs ``out-of-the-box'' to yield state-of-the-art accuracy on the dataset that we tested.  And yet, an expert experimentalist could desire different clustering results, further improving the performance.  Because our inference methodology is based on a biophysical model, all of the hyperparameters have natural and intuitive interpretations.  Therefore, adjusting the performance is relatively intuitive.  
Although all of the results presented above were manifested without any model tuning, we now discuss how one may constitute a single ``knob'' (parameter) that a neuroscientist may ``turn'' to examine different kinds of results.

\begin{figure}[!htbp]
\centering

\subfigure[]{
   \includegraphics[width=0.7\linewidth] {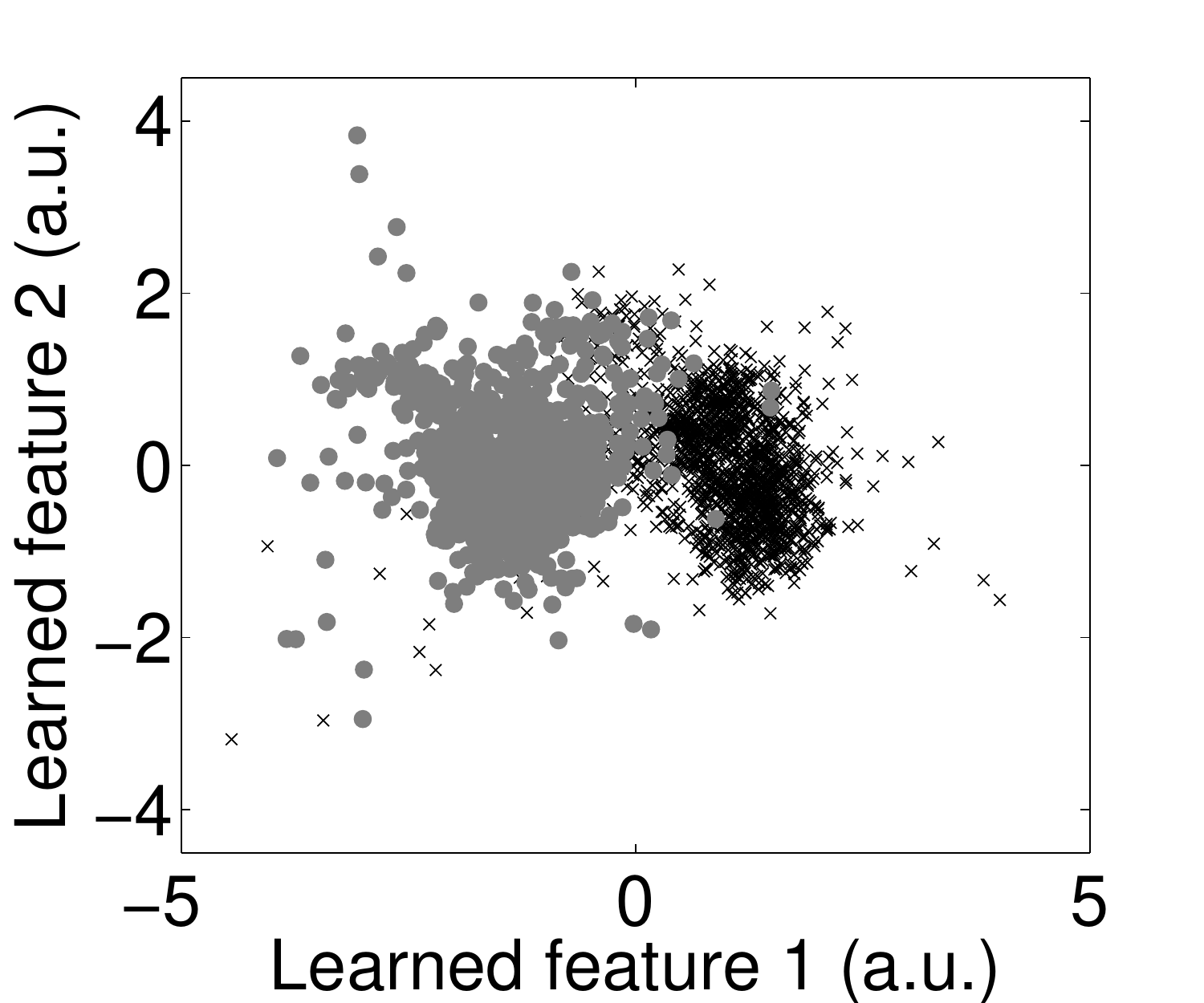}
   \label{fig:Tuning_Parameter1}
 }
 \subfigure[]{
   \includegraphics[width=0.7\linewidth] {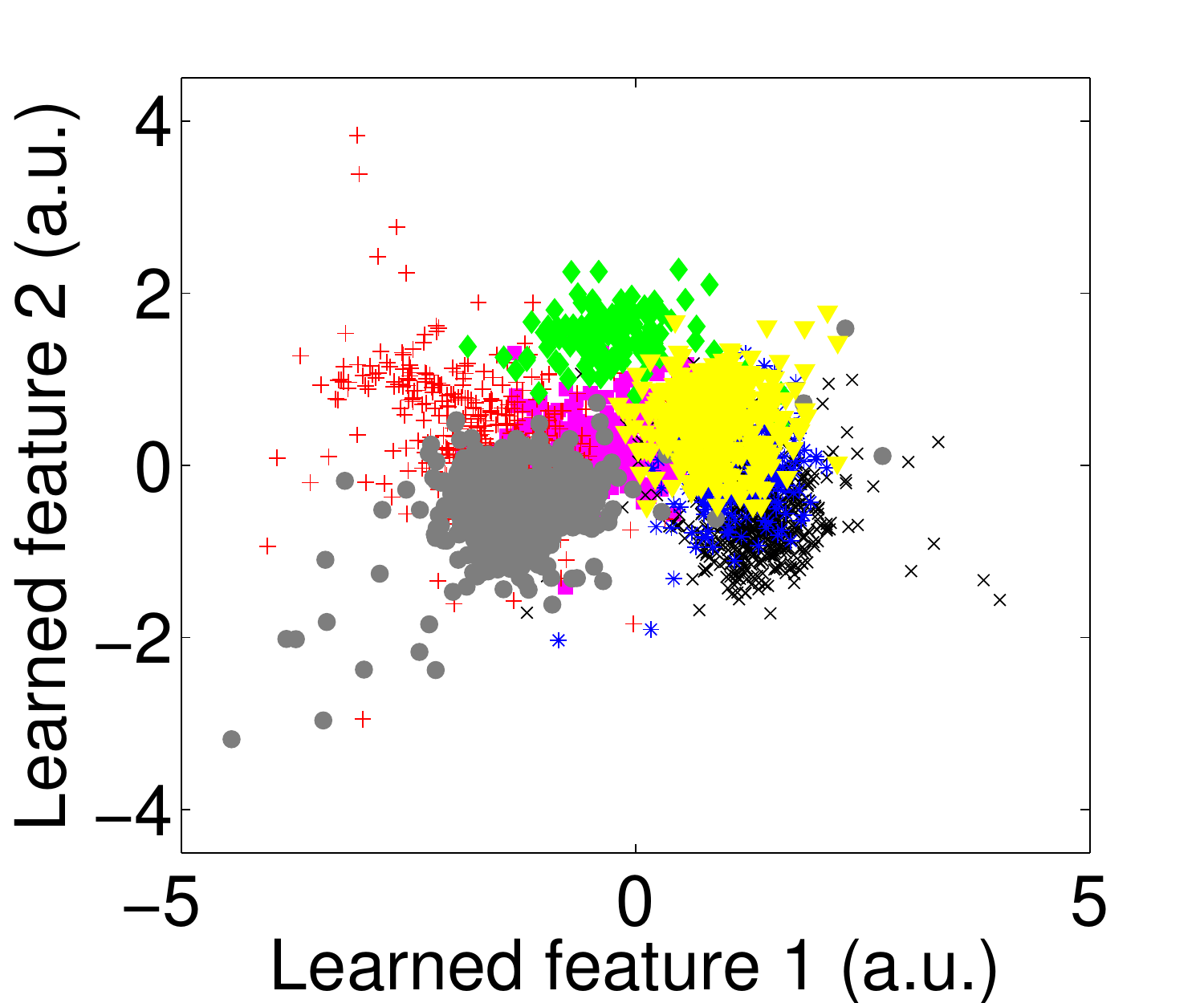}
   \label{fig:Tuning_Parameter2}
 }
  \caption{{
{Effect of manually tuning $\omega_0$ to obtain a different number of features for the rat motor cortex data.}
(a) Waveforms projected down onto two learned features based on cluster result with $\omega_0=10^{6}$, the number of inferred clusters is two. (b) Same as (a) with $\omega_0=10^{8}$; the number of inferred clusters is seven.
   }}\label{fig:Tuning_Parameter} \vspace{-20pt}
\end{figure}

In Section \ref{sec:dict} the variance of additive noise $(e_1,\cdots, e_n)$ are controlled by the covariance $\mbox{diag}(\eta_1^{-1},\cdots, \eta_T^{-1})$. If we set $\mbox{diag}(\eta_1^{-1},\cdots, \eta_T^{-1})=\omega_0^{-1}\Imat_T$, then parameter $\omega_0$ may be tuned to control the variability (diversity) of spikes. The cluster diversity encouraged by setting different values of $\omega_0$ in turn manifests different numbers of clusters, which a neuroscientist may adjust as desired. As an example, we consider the publicly available data from Section \ref{sec:truth}, and clusterings (color coded) are shown for two settings of $\omega_0$ {in } Figure \ref{fig:Tuning_Parameter}. In this figure{,} each spike is depicted in two-dimensional {learned feature} space, taking {two arbitrary features (because features are not inherently ordered)}; this is simply for display purposes, as here feature learning is done via dictionary learning, and in general more than two dictionary components are utilized to represent a given waveform.

The value of $\omega_0$ defines how much of a given signal is associated with noise $\Emat_{ij}$, and how much is attributed to the term $\Dmat\Lambdamat\Smat_{ij}$ characterized by a summation of dictionary elements (see (1)). If $\omega_0$ is large, then the noise contribution to the signal is small (because the noise variance is imposed to be small), and therefore the variability in the observed data is associated with variability in the underlying signal (and that variability is captured via the dictionary elements). Since the clustering is performed on the dictionary usage, if $\omega_0$ is large we expect an increasing number of clusters, with these clusters capturing the greater diversity/variability in the underlying signal. By contrast, if $\omega_0$ is relatively small, more of the signal is attributed to noise $\Emat_{ij}$, and the signal components modeled via the dictionary are less variable (variability is attributed to noise, not signal). Hence, as $\omega_0$ diminishes in size we would expect fewer clusters. This phenomenon is observed in the example in Figure \ref{fig:Tuning_Parameter}, with this representative of behavior we have observed in a large set of experiments{ on the rat motor cortex data}.

\subsection{Sparsely Firing Neurons} \label{sec:sparse}

\begin{figure*}[!htbp]
\centering

   \includegraphics[width=1.0\linewidth]{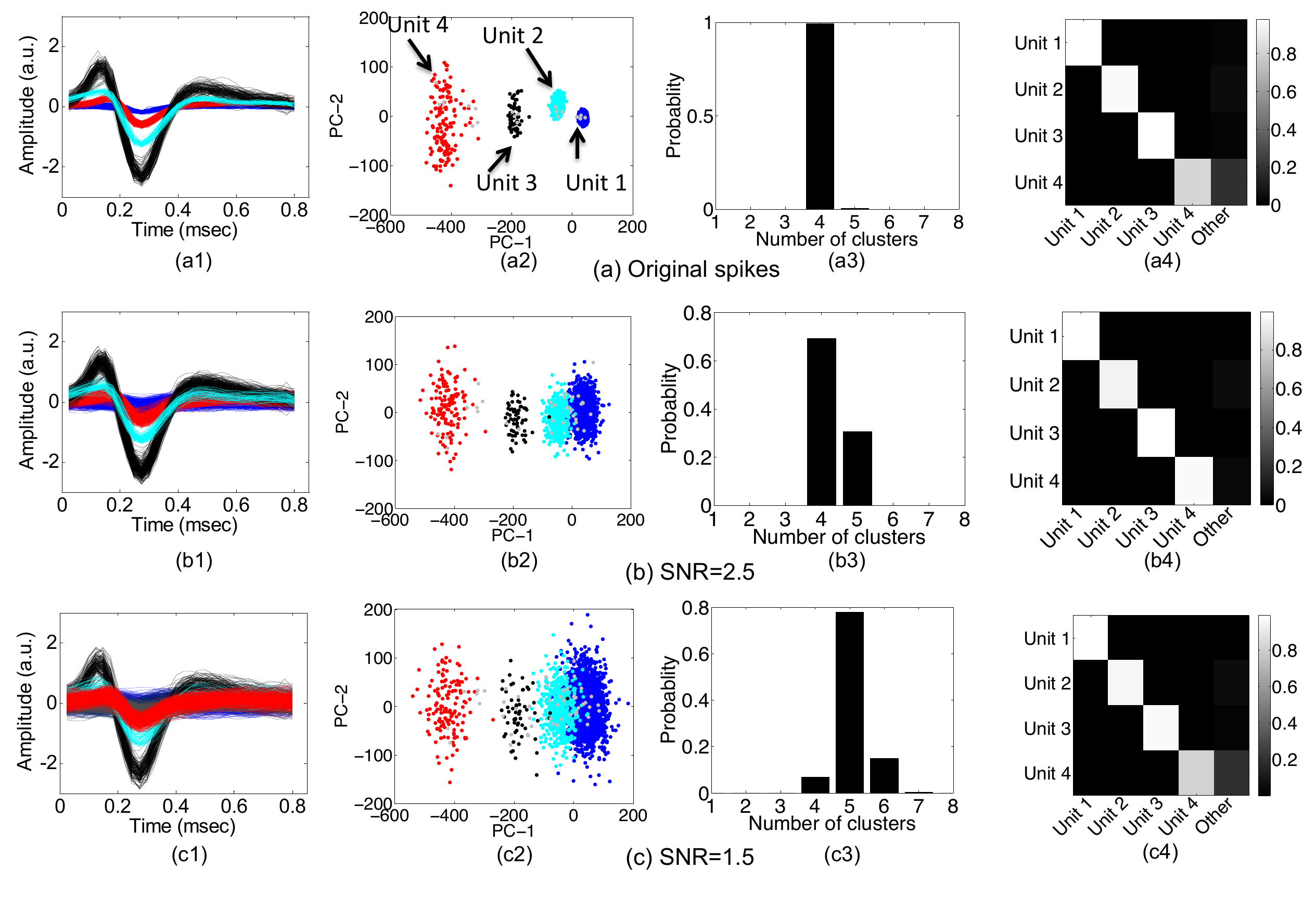}
  \caption{{
Sparse firing results on synthetic data based on the Pittsburgh dataset. 
The three rows correspond to three different signal-to noise ratio (SNR) levels: (a) 1, (b) 1.5, and (c) 2.5. The four columns correspond to: (1) cluster results of spike waveforms with colors representing different clusters, (2) plots of learned features  based on cluster result, (3) approximate posterior distribution of cluster numbers, and (4) confusion matrix heatmap. 
Note that we accurately recover all the sparsely spiking neurons except the sparsest one in the noisiest regime.
   }} \vspace{-10pt}
   \label{fig:Sparse_firing_neuron}
\end{figure*}

Recently, several manuscripts have directly addressed spike sorting in the present of sparsely firing neurons \cite{Pedreira2012, Adamos2012}.  We operationally define a sparsely  firing neuron as a neuron whose spike count has significantly fewer spikes than the other isolated neurons.
Based on reviewer recommendations, we assessed the performance of FMM-DL in such regimes utilizing the following synthetic data.  First, we extracted spike waveforms from four clusters from the new dataset discussed in Section \ref{sub:data_acquisition_and_pre_processing}.  We excluded all waveforms that did not clearly separate (Figure \ref{fig:Sparse_firing_neuron}(a1))
to obtain clear clustering criteria (Figure \ref{fig:Sparse_firing_neuron}(a2)).  There were 2592, 148, 506, and 64 spikes in the first, second, third, and fourth cluster, respectively.
  Then, 
we added real noise---as described in section \ref{sec:eval}---to each waveform at two different levels to obtain increasingly noisy and less-well separated clusters (Figure \ref{fig:Sparse_firing_neuron}(b1), (b2), (c1), and (c2)).  We applied FMM-DL, Wave-clus \cite{Pedreira2012} and Wave-clus ``forced'' (in which we hand tune the parameters to obtain optimal results) and ISOMAP dominant sets \cite{Adamos2012} to all three signal-to-noise ratio (SNR) regimes to assess our relative performance with the following results.

The third column of Figure \ref{fig:Sparse_firing_neuron} shows the posterior estimate of the number of clusters for each of the three scenarios.  As long as SNR is relatively good, for example, higher than 2 in this simulation, the posterior number of clusters inferred by FMM-DL correctly has its maximum at four clusters.  Similarly, for the good and moderate SNR regimes, the confusion matrix is essentially a diagonal matrix, indicating that FMM-DL assigns spikes to the correct cluster.  Only in the poor SNR regime (SNR=1.5), does the posterior move away from the truth.  This occurs because Unit 1 becomes over segmented, as depicted in (c2).  (c4) shows that only this unit struggles with assignment issues, suggestive of the possibility of a post-hoc correction if desired.

\begin{figure*}[!htbp]
\centering
   \includegraphics[width=\textwidth] {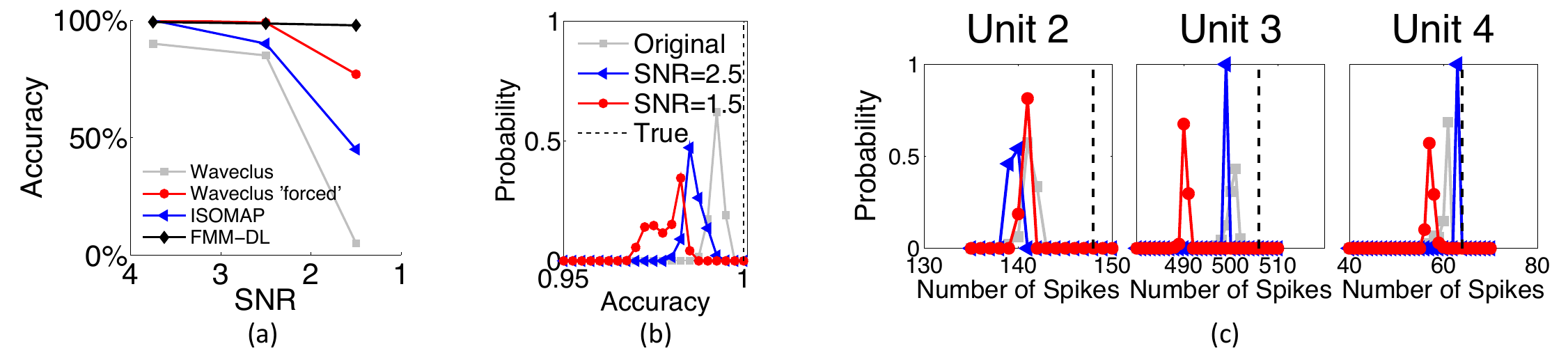}
  \caption{ {Performance analysis in the sparsely firing neuron case on synthetic data based on the Pittsburgh dataset. (a) Accuracy comparisons based on the cluster results under the various SNR. (b) Approximate posterior distributions of error rate for FMM-DL in the different SNR levels. (c) Approximate posterior distributions of spike waveform number for the unit 2, unit 3. and unit 4 under the various SNR regimes. 
   }} \label{fig:posterior_error_rate} \vspace{-10pt}
\end{figure*}

Figure \ref{fig:posterior_error_rate}(a) compares the performance of FMM-DL to previously proposed methods.  Even after fine-tuning the Wave-clus method to obtain its optimal performance on these data, FMM-DL yields a better accuracy.  
In addition to obtaining better point-estimates of spiking, via our Bayesian generative model, we also obtain posteriors over all random variables of our model, including number of spikes per unit.  Figure \ref{fig:posterior_error_rate}(b) and (c) show such posteriors, which may be used by the experimentalist to assess data quality.


\subsection{Computational requirements}

The software used for the tests in this paper were written in (non-optimized) Matlab, and therefore computational efficiency has not been a focus. The principal motivating focus of this study concerned interpretation of {longitudinal} spike waveforms, as discussed in Section \ref{sec:forensics}, for which computation speed is desirable, but there is not a need for real-time processing (for example, for a prosthetic).
 Nevertheless, to give a sense of the computational load for the model, it takes about 20 seconds for each Gibbs sample, when considering analysis of 170800 spikes across $N=8$ channels; computations were performed on a PC, specifically a Lenevo T420 (CPU is Intel(R) Core (TM) i7 M620 with 4 GB RAM). Significant computational acceleration may be manifested by coding in C, and via development of online methods for Bayesian inference (for example, see \cite{Wang11}). In the context of such online Bayesian learning one typically employs approximate variational Bayes inference rather than Gibbs sampling, which typically manifests significant acceleration \cite{Wang11}.

\section{Discussion}\label{sec:conclusions}

\subsection{Summary} 
\label{sub:summary}

A new focused mixture model (FMM) has been developed, motivated by real-world studies with longitudinal electrophysiological data, for which traditional methods like the hierarchical Dirichlet process have proven inadequate. In addition to performing ``focused'' clustering, the model jointly performs feature learning, via dictionary learning, which significantly improves performance over principal components. We explicitly model the count of signals within a recording period {by $p_i$}. The rate of neuron firing constitutes a primary information source \cite{Donoghue07}, and therefore it is desirable that it be modeled. This rate is controlled here by a parameter {$\phi_m^{(i)}$}, and this was allowed to be unique for each recording period $i$.

\subsection{Future Directions} 
\label{sub:future_directions}

In future research one may constitute a mixture model on {$\phi_m^{(i)}$}, with each mixture component reflective of a latent neural (firing) state; one may also explicitly model the time dependence of {$\phi_m^{(i)}$}{, as in the Mixture of Kalmans work} \cite{Calabrese2010}. Inference of this state could be important for decoding neural signals and controlling external devices or muscles. In future work one may also wish to explicitly account for covariates associated with animal activity \cite{Ventura}, which may be linked to the firing rate we model here (we may regress $p_i$ to observed covariates).

In the context of modeling and analyzing electrophysiological data, recent work on clustering models has accounted for refractory-time
violations \cite{Wood2009,Calabrese2010,Bo2011}, which occur when two or more spikes that
are sufficiently proximate are improperly associated with the same
cluster/neuron (which is impossible physiologically due to the refractory time delay
required for the same neuron to re-emit a spike). The methods developed in \cite{Wood2009,Bo2011} may be extended to the class of mixture models developed above. We have not done so for two reasons: ($i$) in the context of everything else that is modeled here (joint feature learning, clustering, and count modeling), the refractory-time-delay issue is a relatively minor issue in practice; and ($ii$) perhaps more importantly, an important issue is that not all components of electrophysiological data are spike related (which are associated with refractory-time issues). As demonstrated in Section \ref{sec:results}, a key component of the proposed method is that it allows us to distinguish single-unit (spike) events from other phenomena.

Perhaps the most important feature of spike sorting methods that we have not explicitly included in this model is ``overlapping spikes'' \cite{Bar-Gad2001, Zhang2004, Wang2006, Vargas-Irwin2007, Herbst2008a, Adamos2010, Franke2010b}. Preliminary analysis of our model in this regime (not shown), inspired by reviewer comments, demonstrated to us that while the FMM-DL as written is insufficient to address this issue, a minor modification to FMM-DL will enable ``demixing'' overlapping spikes.  We are currently pursuing this avenue.  Neuronal bursting---which can change the waveform shape of a neuron---is yet another possible avenue for future work.  

\section*{Acknowledgement} The research reported here was supported by the Defense Advanced Research Projects Agency (DARPA) under the HIST program, managed by Dr. Jack Judy. The findings and opinions in this paper are those of the authors alone.

\section*{Appendix} \label{sec:appendix}

\setcounter{subsection}{0}
\subsection{Connection to Bayesian Nonparametric Models} 
\label{sub:connection_to_previous_bayesian_non_parametrics}


The use of nonparametric Bayesian methods like the Dirichlet process (DP) \cite{Wood2009,Bo2011} removes some of the \emph{ad hoc} character of classical clustering methods, but there are other limitations within the context of electrophysiological data analysis. The DP and related models are characterized by a scale parameter $\alpha>0$, and the number of clusters grows as $\mathcal{O}(\alpha \log S)$ \cite{Teh2010a}, with $S$ the number of data samples. This growth without limit in the number of clusters with increasing data is undesirable in the context of electrophysiological data, for which there are a finite set of processes responsible for the observed data. Further, when jointly performing mixture modeling across multiple tasks, the \emph{hierarchical} Dirichlet process (HDP) \cite{HDP} shares all mixture components, which may undermine inference of subtly different clusters.

%
In this paper we integrate dictionary learning and clustering for analysis of electrophysiological data, as in \cite{Dilan,Bo2011}. However, as an alternative to utilizing a method like DP or HDP \cite{Wood2009,Bo2011} for clustering, we develop a new hierarchical clustering model in which the number of clusters is modeled explicitly; this implies that we model the number of underlying {neurons}---or clusters---separately from the firing rate, with the latter controlling the total number of observations. This is done by integrating the Indian buffet process (IBP) \cite{IBP} with the Dirichlet distribution, similar to \cite{compound}, but with unique characteristics. The IBP is a model that may be used to \emph{learn} features representative of data, and each potential feature is a ``dish'' at a ``buffet''; each data sample (here {a} neuronal spike) selects which features from the ``buffet'' are most appropriate for its representation. The Dirichlet distribution is used for clustering data, and therefore here we jointly perform feature learning and clustering, by integrating the IBP with the Dirichlet distribution. The proposed framework explicitly models the quantity of data (for example, spikes) measured within a given recording interval. {To our knowledge,} this is the first time the firing rate of electrophysiological data is modeled jointly with clustering \emph{and} jointly with feature/dictionary learning. The model demonstrates state-of-the-art clustering performance on publicly available data. Further, concerning distinguishing single-unit-events, we demonstrate how this may be achieved using the {FMM-DL}  method, considering new measured (experimental) electrophysiological data.

\subsection{Relationship to Dirichlet priors} \label{sec:related}

A typical prior for $\piv^{(i)}$ is a symmetric Dirichlet distribution \cite{Dilan},
\beq \piv^{(i)}\sim\mbox{Dir}(\tilde{\alpha}_0/M,\dots,\tilde{\alpha}_0/M).\label{eq:Dir}\eeq
In the limit{,} $M\rightarrow\infty${,} this reduces to a draw from a Dirichlet process \cite{Wood2009,Bo2011}, represented $\piv^{(i)}\sim\mbox{DP}(\tilde{\alpha}_0 G_0)$, with $G_0$ the ``base'' distribution defined in (\ref{eq:mixture}). Rather than drawing each $\piv^{(i)}$ independently {from $\mbox{DP}(\tilde{\alpha}_0 G_0)$}, we may consider the hierarchical Dirichlet process (HDP) \cite{HDP} as
\beq \piv^{(i)}\sim\mbox{DP}(\tilde{\alpha}_1 G)~,~~~~G\sim\mbox{DP}(\tilde{\alpha}_0 G_0)\eeq
The HDP methodology imposes that the $\{\piv^{(i)}\}$ share the same set of ``atoms" $\{\muv_{mn},\Omegamat_{mn}\}$, implying
a sharing of the different types of clusters across the time intervals $i$ at which data are collected. A detailed discussion of the HDP formulation is provided in \cite{Bo2011}.

These models have limitations in that the inferred number of clusters grows with observed data (here the clusters are ideally connected to {neurons}, the number of which will not necessarily grow with  {longer samples}). Further, the above clustering model assumes the number of samples is given, and hence is not modeled (the information-rich firing rate is not modeled).
Below we develop a framework that yields hierarchical clustering like HDP, but the number of clusters and the data count (for example, spike rate) are modeled explicitly.

\subsection{Other Formulations of the FMM} 
\label{sub:other_formulations_of_the_fmm}


Let the total set of data measured during interval $i$ be represented $\bm{\mathcal{D}}_i=\{\Xmat_{ij}\}_{j=1}^{M_i}${, where $M_i$ is the total number of events during interval $i$}. In the experiments below, a ``recording interval'' corresponds to a day on which data were recorded for an hour (data are collected separately on a sequence of days), and the set $\{\Xmat_{ij}\}_{j=1}^{M_i}$ defines all signals that exceeded a threshold during that recording period. In addition to modeling $M_i$, we wish to infer the number of distinct clusters $C_i$ characteristic of $\bm{\mathcal{D}}_i$, and the relative fraction (probability) with which the $M_i$ observations are apportioned to the $C_i$ clusters.

Let $n_{im}^*$ represent the number of data samples in $\bm{\mathcal{D}}_i$ that are apportioned to cluster $m\in\{1,\dots,M\}=\mathcal{S}$, with $M_i$$=\sum_{m=1}^M n_{im}^*$. The set $\mathcal{S}_i\subset \mathcal{S}$, with $C_i=|\mathcal{S}_i|$, defines the \emph{active} set of clusters for representation of $\bm{\mathcal{D}}_i$, and therefore $M$ serves as an upper bound ($n_{im}^*=0$ for $m\in\mathcal{S}\setminus\mathcal{S}_i$).

We impose $n_{im}^*\sim\mbox{Poisson}(b_m^{(i)}\hat{\phi}_m^{(i)})$ with the priors for $b_m^{(i)}$ and $\hat{\phi}_m^{(i)}$ given in Eqs. \eqref{eq:gen1} and \eqref{eq:gen2}.
%
Note that $n_{im}^*=0$ when $b_m^{(i)}=0$, and therefore $\bv^{(i)}=(b_1^{(i)},\dots,b_M^{(i)})^T$ defines indicator variables identifying the active subset of clusters $\mathcal{S}_i$ for representation of $\bm{\mathcal{D}}_i$. Marginalizing out $\hat{\phi}_m^{(i)}$, $n_{im}^*\sim\mbox{NegBin}(b_m^{(i)}{\phi}_m,p_i)$. This emphasize another motivation for the form of the prior: the negative binomial modeling of the counts (firing rate) is more flexible than a Poisson model, as it allows the mean and variance on the number of counts to be different (they are the same for a Poisson model).

While the above methodology yields a generative process for the number, $n_{im}^*$, of elements of $\bm{\mathcal{D}}_i$ apportioned to cluster $m$, it is desirable to explicitly associate each member of $\bm{\mathcal{D}}_i$ with one of the clusters (to know not just \emph{how many} members of $\bm{\mathcal{D}}_i$ are apportioned to a given cluster, but also \emph{which} data are associated with a given cluster). Toward this end, consider the alternative equivalent generative process for $\{n_{im}^*\}_{m=1,M}$ (see Lemma 4.1 in \cite{Mingyuan2012} for a proof of equivalence): first draw
$M_i$$\sim\mbox{Poisson}(\sum_{m=1}^M b_m^{(i)}\hat{\phi}_m^{(i)})$, 
 and then
\beqs & (n_{i1}^*,\dots,n_{iM}^*)\sim\mbox{Mult}(M_i;\pi_1^{(i)},\dots,\pi_M^{(i)})\\ &\pi_m^{(i)}=b_m^{(i)}\hat{\phi}_m^{(i)}/\sum_{m^\prime=1}^M b_{m^\prime}^{(i)}\hat{\phi}_{m^\prime}^{(i)}\label{eq:mixt}\eeqs 
with $\hat{\phi}_m^{(i)}$, $\{{\phi}_m\}$, $\{b_m^{(i)}\}$, and $\{p_i\}$ constituted as in (\ref{eq:gen1})-(\ref{eq:gen2}). Note that we have $M_i$$\sim\mbox{NegBin}(\sum_{m=1}^M b_m^{(i)}{\phi}_m,p_i)$ by marginalizing out $\hat{\phi}_m^{(i)}$.

Rather than drawing $(n_{i1}^*,\dots,n_{iM}^*)\sim\mbox{Mult}(M_i;\pi_1^{(i)},\dots,\pi_M^{(i)})$, for each of the $M_i$ data we may draw indicator variables $z_{ij}\sim\sum_{m=1}^M\pi_m^{(i)}\delta_m$, where $\delta_m$ is a unit measure concentrated at the point $m$. Variable $z_{ij}$ assigns data sample $j\in\{1,\dots,$ $M_i$$\}$ to one of the $M$ possible clusters, and $n_{im}^*=\sum_{j=1}^{M_i} 1(z_{ij}=m)$, with $1(\cdot)$ equal to one if the argument is true, and zero otherwise. The probability vector $\piv^{(i)}$ defined in (\ref{eq:mixt}) is now used within the mixture model in (\ref{eq:mixture}).

As a consequence of the manner in which $\hat{\phi}_m^{(i)}$ is drawn in (\ref{eq:gen1}), and the definition of $\piv^{(i)}$ in (\ref{eq:mixt}), for \emph{any} $p_i\in(0,1)$, the proposed model imposes
\beq \piv^{(i)}\sim\mbox{Dir}(b_1^{(i)}{\phi}_1,\dots,b_M^{(i)}{\phi}_M) \label{eq:gDir} \eeq

\subsection{Additional Connections to Other Bayesian Models} 
\label{sub:additional_connections_to_other_bayesian_models}

Eq. (\ref{eq:gDir}) {demonstrates that } the proposed model is a generalization of (\ref{eq:Dir}). Considering the limit $M\rightarrow\infty$, and upon marginalizing out the $\{\nu_m\}$, the binary vectors $\{\bv^{(i)}\}$ are drawn from the Indian buffet process (IBP), denoted $\bv^{(i)}\sim\mbox{IBP}(\alpha)$. The number of non-zero components in each $\bv^{(i)}$ is drawn from $\mbox{Poisson}(\alpha)$, and therefore for finite $\alpha$ the number of non-zero components in $\bv^{(i)}$ is finite, even when $M\rightarrow\infty$. Consequently $\mbox{Dir}(b_1^{(i)}{\phi}_1,\dots,b_M^{(i)}{\phi}_M)$ is well defined even when $M\rightarrow\infty$ since, with probability one, there are only a finite number of non-zero parameters in $(b_1^{(i)}{\phi}_1,\dots,b_M^{(i)}{\phi}_M)$. This model is closely related to the compound IBP Dirichlet (CID) process developed in \cite{compound}, with the following differences.

Above we have explicitly derived the relationship between the negative binomial distribution and the CID, and with this understanding we recognize the importance of $p_i$; the CID \emph{assumes} $p_i=1/2$, but there is no theoretical justification for this. Note that  $M_i$$\sim\mbox{NegBin}(\sum_{m=1}^M b_m^{(i)}{\phi}_m^{(i)},p_i)$. The mean of $M_i$ is $(\sum_{m=1}^M b_m^{(i)}{\phi}_m) p_i/(1-p_i)$, and the variance is $(\sum_{m=1}^M b_m^{(i)}{\phi}_m)p_i/(1-p_i)^2$. If $p_i$ is fixed to be  {$1/2$} as in \cite{compound}, this implies that we believe that the variance is two times the mean, and the mean and variance of $M_i$ are the same for all intervals $i$ and $i^\prime$ for which $\bv^{(i)}=\bv^{(i^\prime)}$. However, in the context of electrophysiological data, the rate at which neurons fire plays an important role in information content \cite{Donoghue07}. Therefore, there are many cases for which intervals $i$ and $i^\prime$ may be characterized by firing of the same neurons ($i.e.$, $\bv^{(i)}=\bv^{(i^\prime)}$) but with very different rates ($M_i\neq M_{i^\prime}$). The modeling flexibility imposed by inferring $p_i$ therefore plays an important practical role for modeling electrophysiological data, and likely for other clustering problems of this type.

To make a connection between the proposed model and the HDP, motivated by (\ref{eq:gen1})-(\ref{eq:gen2}), consider $\bar{\phiv}=(\bar{\phi}_1,\cdots,\bar{\phi}_M) \sim \mbox{Dir}(\gamma_0,\cdots,\gamma_0)$, which corresponds to $(\phi_1,\dots,\phi_M)/\sum_{m^\prime=1}^M \phi_{m^\prime}$. From $\bar{\phiv}$ we yield a \emph{normalized} form of the vector $\phiv=(\phi_1,\dots,\phi_M)$. The normalization constant $\sum_{m=1}^M\phi_m$ is lost after drawing $\bar{\phiv}$; however, because $\phi_m\sim\mbox{Ga}(\gamma_0,1)$, we may consider drawing $\tilde{\alpha}_1\sim\mbox{Ga}(M\gamma_0,1)$, and \emph{approximating} ${\phiv}\approx\tilde{\alpha}_1\bar{\phiv}$. With this approximation for $\phiv$, $\piv^{(i)}$ may be drawn approximately as $\piv^{(i)}\sim\mbox{Dir}(\tilde{\alpha}_1b_1^{(i)}\bar{\phi}_1,\dots,\tilde{\alpha}_1b_M^{(i)}\bar{\phi}_M)$. This yields a simplified and approximate hierarchy
\beqs & \piv^{(i)}\sim\mbox{Dir}(\tilde{\alpha}_1(\bv^{(i)}\odot\bar{\phiv}))\\ &\bar{\phiv}=(\bar{\phi}_1,\cdots,\bar{\phi}_M) \sim \mbox{Dir}(\gamma_0,\cdots,\gamma_0),~\tilde{\alpha}_1\sim\mbox{Ga}(M\gamma_0,1)\nonumber\eeqs
with $\bv^{(i)}\sim\mbox{IBP}(\alpha)$ and $\odot$ representing a pointwise/Hadamard product. If we consider $\gamma_0=\hat{\alpha}_0/M$, and the limit $M\rightarrow\infty$, with $\bv^{(i)}$ all ones, this corresponds to the HDP, with $\hat{\alpha}_1\sim\mbox{Ga}(\hat{\alpha}_0,1)$. 
{We call such a model the non-focused mixture model (NFMM).}
Therefore, the proposed model is intimately related to the HDP, with three differences: ($i$) $p_i$ is not restricted to be 1/2, which adds flexibility when modeling counts; ($ii$) rather than drawing $\bar{\phiv}$ and the normalization constant $\tilde{\alpha}_1$ separately, as in the HDP, in the proposed model $\phiv$ is drawn directly via $\phi_m\sim\mbox{Ga}(\gamma_0,1)$, with an explicit link to the count of observations $M_i\sim\mbox{NegBin}(\sum_{m=1}^Mb_m^{(i)}\phi_m,p_i)$; and ($iii$) the binary vectors $\bv^{(i)}$ ``focus'' the model on a sparse subset of the mixture components, while in general, within the HDP, all mixture components have non-zero probability of occurrence for all tasks $i$. As demonstrated in Section \ref{sec:results}, this focusing nature of the proposed model is important in the context of electrophysiological data.


\subsection{Proof of Lemma 3.1}

\begin{proof}
Denote $w_{j}= \sum_{l=1}^ju_{l}$, $j=1,\cdots,m$. Since $w_{j}$ is the summation of $j$ iid $\mbox{Log}(p)$ distributed random variables, the probability generating function of $w_{j}$ can be expressed as
$
G_{W_{j}}(z)=
\left[{\ln(1-pz)}/{\ln(1-p)}\right]^j,~ |z|<{p^{-1}}
$, thus we have
\beqs
& \mbox{Pr}(w_{j}=m) = {G_{W_{j}}^{(m)}(0)}/{m!} = \frac{d^{m}}{dz^{m}} [\ln(1-p z)/{\ln(1-p)}]^j \nonumber\\ &= (-1)^m p^j j! s(m,j)/[\ln(1-p)]^j \eeqs
where we use the property that $[\ln(1+x)]^j = j!\sum_{n=j}^\infty\frac{s(n,j)x^n}{n!}$ \cite{johnson2005univariate}.
Therefore, we have
\beqs
& \mbox{Pr}(\ell = j|-) \propto  \mbox{Pr}(w_{j}=n) \mbox{Pois}(j;-r\ln(1-p)) \nonumber\\ &\propto (-1)^{n+j}s(n,j)/n! r^j =F(n,j)r^j. \qedhere
\eeqs
\end{proof}
The values $F(n,j)$ can be iteratively calculated and each row sums to one, e.g.,
the 3rd to 5th rows are
\[ \left( \begin{array}{ccccccc}
2/3! & 3/3! & 1/3! & 0 & 0 & 0& \cdots \\
6/4! & 11/4! & 6/4! & 1/4! & 0 & 0& \cdots \\
24/5! & 50/5! & 35/5! & 10/5! & 1/5! & 0& \cdots \\
\end{array} \right).\]
To ensure numerical stability when $\phi>1$, we may also iteratively calculate the values of $R_\phi(n,j)$.

\small\bibliography{combined}
\bibliographystyle{plain}



\end{document}